\documentclass[iop]{emulateapj}
\usepackage{epsf}
\usepackage{multirow}
\usepackage{url}
\bibpunct{(}{)}{;}{a}{}{,}
\usepackage{color}
\usepackage{listings}

\newcommand{\hmpc}{\,h^{-1}\,{\rm Mpc}}
\newcommand{\hmpcc}{\,h^{3}\,{\rm Mpc}^{-3}}

\begin{document}
\title{The Impact of Assembly Bias on the Galaxy Content of Dark Matter 
Halos}
\author{Idit Zehavi\altaffilmark{1,*}, Sergio Contreras\altaffilmark{2,3,*},
Nelson Padilla\altaffilmark{2,4}, Nicholas J.\ Smith\altaffilmark{1,5}, 
Carlton M.\ Baugh\altaffilmark{3}, \\ and Peder Norberg\altaffilmark{3,6}}

\altaffiltext{*}{Email: idit.zehavi@case.edu (IZ); stcontre@uc.cl (SC)}
\altaffiltext{1}{Department of Astronomy, Case Western Reserve University,
10900 Euclid Avenue, Cleveland, OH 44106, USA}
\altaffiltext{2}{Instit\'uto Astrof\'{\i}sica, Pontifica Universidad Cat\'olica
de Chile, Santiago, Chile}
\altaffiltext{3}{Institute for Computational Cosmology, Department of Physics, 
Durham University, South Road, Durham, DH1 3LE, UK}
\altaffiltext{4}{Centro de Astro-Ingenier\'{\i}a, Pontificia 
 Universidad Cat\'olica de Chile, Santiago, Chile}
\altaffiltext{5}{Department of Astronomy, Indiana University, 727 E.\ Third
Street, Bloomington, IN 47405, USA}
\altaffiltext{6}{Centre for Extragalactic Astronomy, Department of Physics, 
Durham University, South Road, Durham, DH1 3LE, UK}

\begin{abstract}
We study the dependence of the galaxy content of dark matter halos on 
large-scale environment and halo formation time using semi-analytic galaxy
models applied to the Millennium simulation.  We analyze  subsamples of 
halos at the extremes of these distributions and measure the occupation 
functions for the galaxies they host. We find distinct differences in these 
occupation functions.  The main effect with environment is that central 
galaxies (and in one model also the satellites) in denser regions start 
populating lower-mass halos. A similar, but significantly stronger, trend 
exists with halo age, where early-forming halos are more likely to host 
central galaxies at lower halo mass. We discuss the origin of these trends 
and the connection to the stellar mass -- halo mass relation. We find that, 
at fixed halo mass, older halos and to some extent also halos in dense 
environments tend to host more massive galaxies. Additionally, we see a 
reverse trend for the satellite galaxies occupation where early-forming 
halos have fewer satellites, likely due to having more time for them to 
merge with the central galaxy. We describe these occupancy variations 
also in terms of the changes in the occupation function parameters, which can
aid in constructing realistic mock galaxy catalogs.  Finally, we study the 
corresponding galaxy auto- and cross-correlation functions of the different 
samples and elucidate the impact of assembly bias on galaxy clustering. Our 
results can inform theoretical models of assembly bias and attempts to detect 
it in the real universe. 
\end{abstract} 


\section{Introduction}
\label{Sec:Intro}

In the standard paradigm of hierarchical structure formation, galaxies reside 
inside dark matter halos.  The formation and evolution of these halos is 
dominated by gravity and can be well predicted using high-resolution numerical
simulations and in some cases analytic models.  The formation of the galaxies 
and their relation to the dark matter halos is more complex and depends on 
the detailed physical processes leading to the varied observed galaxy 
properties.

It has been well established that the
local halo environment of galaxies plays a fundamental role in shaping 
their properties. In particular, local effects are thought to be 
responsible for the transformation of blue, late-type and star-forming 
galaxies into red, early-type and passive galaxies (see, e.g., 
\citealt{oemler74,dressler80,lewis02,balogh04b,baldry04,blanton09}),
even though there is no consensus on the relative importance of the 
specific processes that play a role. Different mechanisms such as mergers 
and interactions, ram-pressure stripping of cold gas, 
starvation or strangulation and harassment all lead to changes in galaxies
morphologies within the host halo environment. It is not clear, however, 
to what extent are galaxy properties affected by their overall ``global'' 
environment on scales larger than the individual halos. While there is 
evidence that global environments affect galaxy populations ---
for example, red galaxies frequent high-density environments while blue 
galaxies are prevalent in low-density regions (e.g., 
\citealt{hogg03,blanton05,blanton06,blanton09}) --- it is debatable whether 
the large-scale environment has an actual impact on the physical processes 
involved in galaxy formation and evolution. 

A useful approach for studying the predictions of galaxy formation processes 
is with semi-analytic modeling (SAM) of galaxy formation, in which halos 
identified in large N-body simulations are populated with galaxies and evolved 
according to specified prescriptions for gas cooling, gas formation, feedback 
effects and merging (e.g., \citealt{Cole00,Baugh06,Croton06}). 
These models have been successful in reproducing several measured properties 
such as the galaxy luminosity and stellar mass functions  (see e.g., 
\citealt{Bower06,Guo11,Guo13Sams,Lacey16}).
An alternative way of studying galaxy formation is using hydrodynamic 
simulations which follow the physical baryonic processes by a combination 
of solving the fluid equations and sub-grid prescriptions (see, e.g., 
\citealt{somerville15,Guo16}). Cosmological hydrodynamical simulations are 
starting to play a major role in the study of galaxy formation and evolution. 
Comparisons of such simulations with observations show broad agreement 
(e.g., \citealt{Vogelsberger14,Schaye15,Artale16}).

A popular approach to empirically interpret observed galaxy clustering 
measurements as well as to characterize the predictions of galaxy formation
theories is the Halo Occupation Distribution (HOD) framework (e.g., 
\citealt{peacock00,seljak00,scoccimarro01,berlind02,cooray02,zheng05}). The HOD 
formalism characterizes the relationship between galaxies and dark matter halos
in terms of the probability distribution, $P(N|M_{\rm h})$, that a halo of 
virial mass $M_{\rm h}$ contains $N$ galaxies of a given type, together with 
the spatial and velocity distribution of galaxies inside halos. The 
fundamental ingredient of the modeling is the halo occupation function, 
$\langle N(M_{\rm h}) \rangle$, which represents the average number of galaxies 
as a function of halo mass. The typically assumed shape for the halo occupation
function is motivated by predictions of hydrodynamic simulations and 
semi-analytic models (e.g, \citealt{zheng05}). 
It is often useful to consider separately the contributions from the 
central galaxies, namely the main galaxy at the center of the halo, 
and that of the additional satellite galaxies that populate the halo
\citep{kravtsov04,zheng05}. 
The HOD approach has been demonstrated to be a powerful theoretical tool to
study the galaxy-halo connection, effectively transforming measurements 
of galaxy clustering into a physical relation between galaxies and dark matter
halos. This approach has been very successful in explaining 
the shape of the galaxy correlation function, its environment dependence
and overall dependence on galaxy properties (e.g., 
\citealt{zehavi04,zehavi05b,zehavi11,berlind05,abbas06,skibba06,tinker08,coupon12}).

A central assumption in the conventional applications of this framework is that 
the galaxy content in halos only depends on halo mass and is statistically 
independent of the halo's larger scale environment. This assumption has its
origins in the uncorrelated nature of random walks describing halo assembly in
the standard implementations of the excursion set formalism which results 
in the halo environment being correlated with halo mass, but uncorrelated 
with formation history at fixed mass \citep{bond91,white99,lemson99}.
In this picture, the change in the fraction of blue and red galaxies in 
different large-scale environments, for example, is fully derived from the 
change in the halo mass function in these environments \citep{mo96,lemson99}. 
Consequently, it is not evident that global environments play a major role 
in directly shaping galaxy properties and in particular the HOD.  

This ansatz has been challenged in the last
decade by the demonstration in simulations that the clustering of halos of 
fixed mass varies with halo formation time, concentration and substructure
occupation
\citep{sheth04,gao05,gao07,Jing06,Harker06,Wechsler06,Wetzel06,Angulo08,Pujol14,Lazeyras17}. The dependence of halo clustering on properties other 
than the halo mass has broadly been referred to as {\it halo assembly bias}.
The dependences on the various halo properties manifest themselves in 
different ways and are not trivially derived from the correlation between
these properties (see, e.g., \citealt{mao17}).  While a prediction of 
$\Lambda$CDM, the exact physical origin of assembly bias remains unclear, 
but different explanations have been put forth, such as 
correlated modes which break down the random walk assumption, statistics of
peaks and truncation of low-mass halo growth in dense environments 
\citep{keselman07,sandvik07,zentner07,dalal08,desjacques08,hahn09,ludlow11,lacerna11,zhang14,borz16}.

A current topic of active debate is to what extent are galaxies affected 
by the assembly history of their host halos. The stochasticity in the
complex baryonic physics may act to erase the record of halo assembly history. 
If, however, the galaxy properties closely
correlate with the halo formation history, this would lead to a dependence
of the galaxy content on large-scale environment and a corresponding clustering
signature. This effect has commonly been referred to as {\it galaxy assembly 
bias} both colloquially and in the literature, and we adopt this distinction 
here. We stress, however, that what is referred to here is the 
manifestation of halo assembly bias in the galaxy distribution.
The predictions for galaxy assembly bias have been explored with simulated 
galaxies \citep{croton07,reed07,Zhu06,Zu08,zentner14,angulo16,bray16}.  
Detecting (galaxy) assembly bias is much more challenging since halo 
properties are not directly observed.
Observational studies of assembly bias have generally produced mixed results.
There have been several suggestive detections in observations \citep{yang06,berlind06,wang08,cooper10,tinker12,wang13b,lacerna14b,watson15,Hearin15,miyatake16,saito16,zentner16,montero17,tinker17b}  while numerous other studies indicate the impact of assembly bias to be small \citep{abbas06,blanton07,croton08,tinker08,tinker11,deason13,lacerna14a,lin16,ZuMan16,vakili16,dvornik17}.
The situation is further complicated as various systematic effects can mimic
the effects of assembly bias 
(e.g., \citealt{campbell15,Zu16,Zu17,Busch17,Sin17,tinker17a,lacerna17}) and 
the evidence for assembly bias to date remains inconclusive and controversial. 

Such galaxy assembly bias, if significant, would have direct implications 
for interpreting galaxy clustering using the HOD framework (e.g., 
\citealt{Pujol14,zentner14}), as secondary halo parameters in addition to 
the mass or, more generally, the large-scale environment in which the halo 
formed,  would also impact the halo occupation function.  For clarity, we
term these variations of the halo occupation functions as {\it occupancy
variation}.  These effects are all directly related of course, as it is 
exactly this occupancy variation coupled with halo assembly bias that gives 
rise to galaxy assembly bias.

In this paper, we aim to gain further insight and clarify this important 
topic by exploring explicitly the dependence of the halo occupation functions
on the large-scale environment and formation redshift in semi-analytic models.
Limited work has been carried out in directly studying the environmental 
dependencies of the HOD of galaxies, with varied results.
Different works examined the dependence of the subhalo occupation on 
age (e.g., \citealt{jiang16}) and environment \citep{Croft12}, that can
be regarded as a proxy of the satellite occupation, if neglecting the 
effects of baryons. 
\citet{Zhu06} explore the age dependence of the conditional luminosity 
function in a semi-analytic model and a hydrodynamical simulation. 
\citet{berlind03} and \citet{mehta14} explore the variations of the HOD in
cosmological hydrodynamical simulations finding no detected dependence on 
environment. \citet{mcewen16} have recently investigated this using 
the age-matching mock catalogs of \citet{Hearin13} (which by design 
exhibit significant assembly bias) detecting a dependence of the HOD on 
environment, mostly for the central galaxy occupation function. 
While the impact of assembly bias on galaxy clustering has already been
demonstrated using a SAM applied to the Millennium simulation 
\citep{croton07,Zu08}, the variation of the HOD itself with large-scale 
environment or other halo properties has not been explored for it.

Here, we use the HOD formalism to directly study the impact of galaxy
assembly bias as predicted by SAMs. We use the output of two independently
developed SAMs, from the Munich and Durham groups, at different number
densities. We measure the halo occupation functions for different large-scale
environment regimes as well as for different ranges of halo formation redshift.
This allows us to assess which features of the HODs vary with environment
and halo age, and we present the corresponding changes in the HOD parameters. 
Additionally, we investigate the galaxy cross-correlation functions for these
different regimes, which highlights the impact of assembly bias on
clustering. Such studies will inform theoretical models incorporating assembly 
bias into halo models as well as attempts to determine it in observational 
data. Additionally, it can facilitate the creation of mock catalogs 
incorporating this effect.

The outline of the paper is as follows.  In Section~\ref{Sec:GFM} we describe
the galaxy formation models used.  In Section~\ref{Sec:HOD} we explore the
dependence of the HOD on large-scale environment and halo age. We discuss
the origin of the trends and the connection to the stellar mass -- halo mass
relation in Section~\ref{Sec:SMHM}. In 
Section~\ref{Sec:clustering} we investigate the clustering dependence on these
properties and we conclude in Section~\ref{Sec:summary}.  
Appendix~\ref{Sec:Cuts} shows our halo-mass dependent sample cuts, while
Appendix~\ref{Sec:AutoCorr} presents further measurements of the 
auto-correlation functions.

\vspace{0.2cm}
\section{The galaxy formation models} 
\label{Sec:GFM}

\vspace{0.1cm}
\subsection{Semi-analytic models}
\label{SubSec:SAM}

The SAMs used in our work are those of \citet{Guo11} (hereafter G11) and 
\cite{Lagos12} (hereafter L12) \footnote{The G11 and L12 outputs are publicly 
available from the Millennium Archive in Garching 
\url{http://gavo.mpa-garching.mpg.de/Millennium/} and Durham
\url{http://virgodb.dur.ac.uk/}}.
The objective of SAMs is to model the main physical processes 
involved in galaxy formation and evolution in a cosmological context: 
(i) the collapse and merging of dark matter halos; 
(ii) the shock heating and radiative cooling of gas inside dark matter 
halos, leading to the formation of galaxy discs; (iii) quiescent star 
formation in galaxy discs; (iv) feedback from supernovae (SNe), from accretion 
of mass onto supermassive black holes and from photoionization heating of 
the intergalactic medium (IGM); (v) chemical enrichment of the stars and gas; 
(vi) dynamically unstable discs; (vii) galaxy mergers driven by dynamical 
friction within dark matter halos, leading to the formation of stellar 
spheroids, which may also trigger bursts of star formation. 
The two models have different implementations of each of these processes. 
By comparing models from different groups we can get a sense for 
which predictions are robust and which depend on the particular 
implementation of the galaxy formation physics (e.g., \citealt{Contreras13}).

The G11 model is a version of {\tt L-GALAXIES}, the SAM code of the Munich
group and is an updated version of earlier implementations
\citep{Delucia04,Croton06,Delucia07}. The L12 model is a development of the 
{\tt GALFORM} Durham model \citep{Bower06,Font08}, which includes an improved 
treatment of star formation, separating the interstellar 
medium into molecular and atomic hydrogen components \citep{Lagos11}. 
An important difference between G11 and L12 is the treatment of
satellite galaxies. In L12, a galaxy is assumed to be stripped of
its hot gas halo completely once it becomes a satellite and start 
decaying onto the central galaxy. In G11, these processes are more
gradual and depend on the destruction of the subhalo and the orbit of 
the satellite. 

\vspace{0.1cm}
\subsection{N-Body simulation and halo merger trees}
\label{SubSec:NBody}

The SAMs used here are both implemented in the Millennium simulation 
\citep{Springel05}. This simulation has a periodic volume of $(500 \hmpc)^3$ and
contains $2160^3$ particle with a mass of $8.61 \times 10^8 M_{\odot}/h$ each.
The simulation has 63 snapshots between $z=127$ and $z=0$ and was run with a 
$\rm \Lambda CDM$ cosmology\footnote{The values of the 
cosmological parameters used in the Millennium simulation are: 
$\Omega_{\rm b}$ =0.045, $\Omega_{\rm M}$ = 0.25, $\Omega_{\Lambda}$ = 0.75, 
h = $H_0/100$ = 0.73, $n_{\rm s}$ = 1, $\sigma_8$ = 0.9.}. 
The G11 and L12 models both use a friends-of-friends 
({\tt FoF}) group finding algorithm \citep{Davis85} to identify halos in each 
snapshot of the simulation, retaining those with at least 20 particles. 
{\tt SUBFIND} is then run on these groups to identify subhalos 
\citep{Springel01}. The merger trees differ from this point on. 
G11 construct dark matter halo merger trees by linking a subhalo 
in one output to a single descendant subhalo in the subsequent snapshot. 
The halo merger tree used in {\tt L-GALAXIES} is therefore a subhalo merger 
tree. L12 employ the {\tt Dhalo} merger tree construction
(\citealt{Jiang14}; see also \citealt{Merson13}) that also uses 
the outputs of the {\tt FoF} and {\tt SUBFIND} algorithms. 
The {\tt Dhalo} algorithm applies conditions on the amount of mass 
stripped from a subhalo and its distance from the center of the 
main halo before it is considered to be merged with the main subhalo. 
Subsequent output times are examined to see if the subhalo moves 
away from the main subhalo, to avoid merging subhalos which have merely 
experienced a close encounter before moving apart. {\tt GALFORM} 
post-processes the {\tt Dhalo} trees to ensure that the halo mass increases 
monotonically with time. 

Consequently, the definition of halo mass used in the two models is not 
the same. The {\tt Dhalo} mass used in {\tt GALFORM} corresponds to an integer 
number of particle masses whereas a virial mass is calculated in 
{\tt L-GALAXIES}. This leads to slight differences in the halo mass function 
between the models. In previous works that focused on comparing the HODs of 
the different models (e.g., \citealt{Contreras17}), we had matched the halo
mass definitions.  Here, since it is not our aim to compare the HODs themselves
in detail, but rather examine the environmental effects on each, 
we prefer to leave the halo mass definitions as is, but we point out that 
some differences between the models are due to this. A comparison of 
{\tt Dhalo} masses and other halo definitions is presented in \citet{Jiang14}.

\vspace{0.2cm}
\section{The HOD dependence on environment and halo age}
\label{Sec:HOD}
A fundamental assumption of the HOD approach is that the galaxy content in
halos depends only on the mass of the host halo. Any dependence of the HOD on 
secondary parameters, like halo age or large-scale environment, is a direct
reflection of galaxy assembly bias (as discussed in \S~\ref{Sec:Intro}).
In this section we examine the impact of halo age and environment on the HOD,
as predicted in the SAMs. In \S~\ref{SubSec:samples} we provide details on 
how the halo age and large-scale environment are defined and their relation to
one another. 
Our main results regarding how the halo occupation functions vary with 
environment and halo age are presented in \S~\ref{SubSec:HOD}, additional
cases are studied in \S~\ref{SubSec:otherHODs}, and the impact
on HOD parameters is shown in \S~\ref{SubSec:param}.

\vspace{0.1cm}
\subsection{Halo formation time and environment}
\label{SubSec:samples}
We define the formation redshift of a halo, as is commonly done, as the 
redshift when the main progenitor reached (for the first time) half of its 
present-day mass.  We obtain this by following the halo merger trees of the 
different models 
and linearly interpolating between the time snapshots available.
For defining the large-scale environment of the halos we use the density 
field obtained directly from the dark-matter particle distribution with a 
$5 \hmpc$ Gaussian smoothing (which we denote as $\delta_5$).  
This was calculated in cells of $\sim 2 \hmpc$ and is provided in the 
database. The $5 \hmpc$ smoothing scale is chosen as it is 
significantly larger than the size of the largest halos, so as to reflect 
the large-scale environment,  and yet have enough different environments
sampled. We test also the other smoothing radii provided in the database,  
$1.25$, $2.5$ and 
$10 \hmpc$, finding the same qualitative trends we find with $5 \hmpc$ for all 
the results shown in this paper.  Alternative density and environment 
definitions are explored in the literature (e.g., \citealt{muldrew11}).  
Observationally, one naturally 
must resort to using the galaxy distribution to define the environment. Here,
as it is available,  we prefer to directly use the underlying dark matter 
density field, though in practice we expect
our results to be insensitive to the details of the definitions.

\begin{figure}

\vspace{-0.4cm}
\hspace{-0.6cm}
\includegraphics[width=0.53\textwidth]{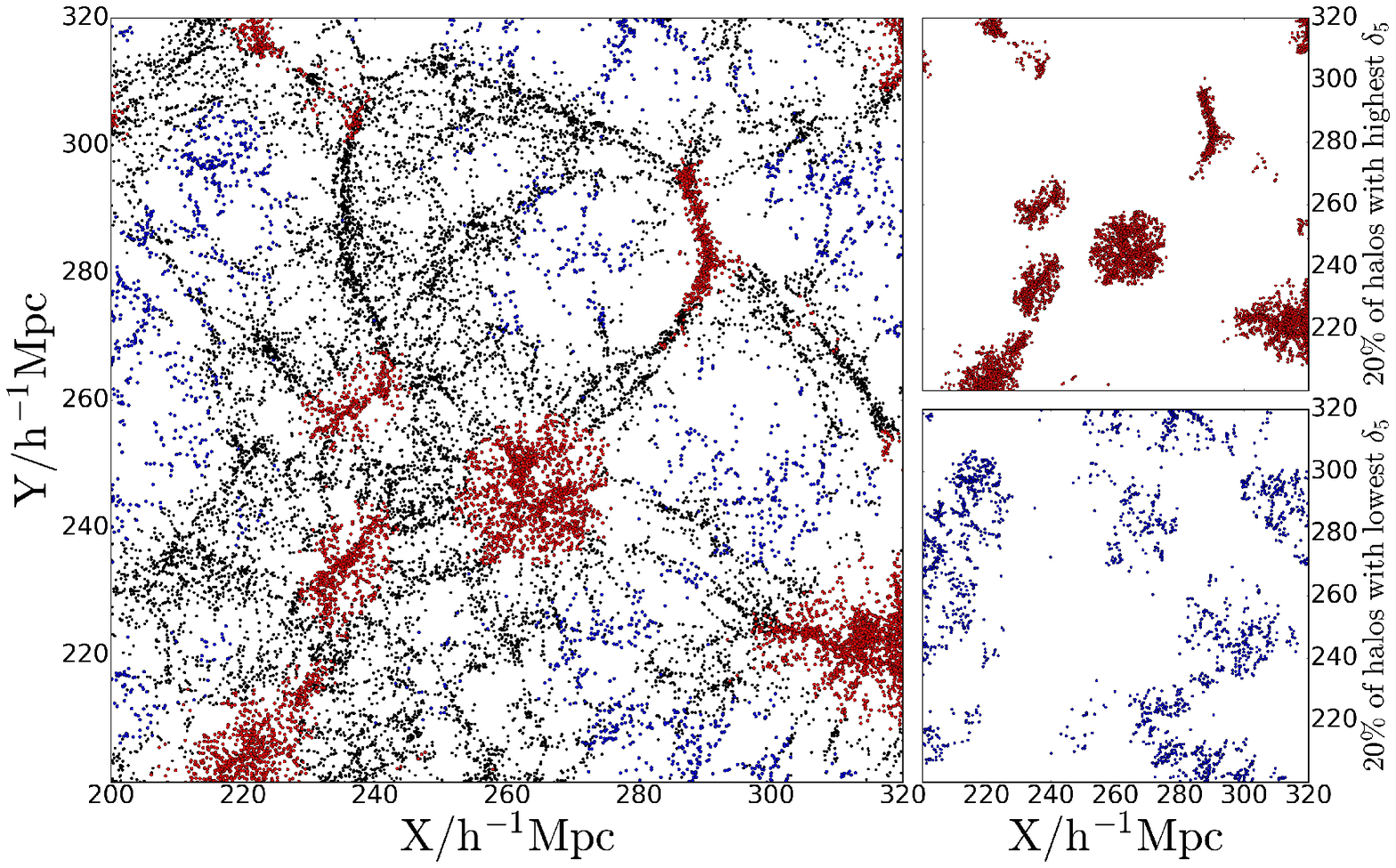}

\hspace{-0.6cm}
\includegraphics[width=0.53\textwidth]{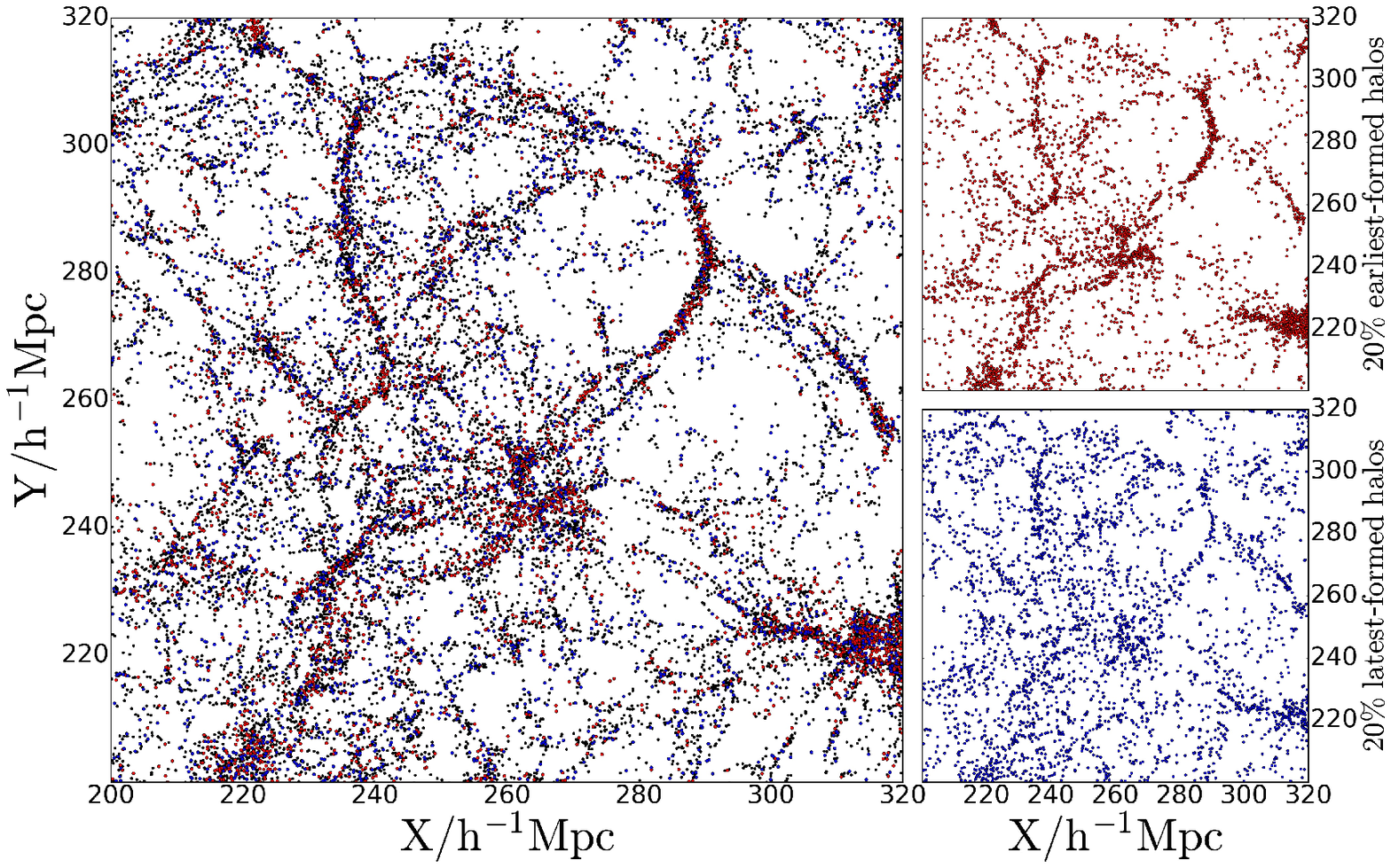}
\caption[]{\label{Fig:CW}
(Top panels) A $120 \hmpc$ x $120 \hmpc$ x $20 \hmpc$ slice of the 
Millennium simulation showing the distribution of the halos in it.
Red (blue) dots represent the 20\% of halos that live in the densest 
(least dense) environments, and the remainder are represented as black dots. 
The density selection is made in 0.2 dex bins of fixed halo mass (see text). 
The bigger plot on the left includes all halos, while the smaller ones on the 
right hand side show separately only the 20\% of halos that live in the 
densest and least dense environments.
(Bottom panels) Same as in the top panels, for the identical slice from the
Millennium simulation, but now color-coding halos by formation time instead 
of environment. Red (blue) dots represent the 20\% most early (late) formed
halos.
}
\end{figure}

To classify the halos by environment, we rank the halos by density in narrow 
(0.2 dex) bins of halo mass and select in each bin the 20\% of halos that 
are in the densest environment and the 20\% of halos in the least dense
environment. This factors out the dependence of the 
halo mass function on environment and allows us to compare the HODs in the 
different environments for halos of nearly equal mass. We follow a similar 
procedure to select the 20\% of halos with the highest and lowest 
formation redshifts. We illustrate how our environment and halo age cuts 
vary with halo mass in Appendix~\ref{Sec:Cuts}. 
We have verified that our mass bins are sufficiently small by using also 
0.1 dex bins and confirming that our results do not change.
We also test splitting the sample into the 10\% and the 50\% extremes of the 
population, and find similar trends as found for the 20\% subsamples.

The distribution of halos classified as residing in the 20\% most 
and least dense environments is shown using 
red and blue dots, respectively, in the top panels of Fig.~\ref{Fig:CW}, 
for a slice from the Millennium simulation. The remainder of the halos are 
shown as black dots. 
The dense and under dense regions appear to ``carve out'' disjoint regions in 
the cosmic web of structure, with the densest ones being more compact than the
underdense regions, as can be expected. The corresponding classification for 
the early- and late-forming halos, for the same slice, is shown in the 
bottom panels of Fig.~\ref{Fig:CW}. 
It is apparent that the distribution of early- and late-forming halos is 
distinctly different than that of halos in dense and underdense environments.
There is perhaps a tendency for the early-forming halos to preferentially 
occupy the dense environments, and a slight trend for late-forming halos to 
populate also the underdense regions. However, the general distribution is very 
different with both early- and late-forming  halos tracing well the cosmic 
web, in contrast to the strong environment patchy pattern. It is also 
clear, even by visual inspection, that the early-forming halos are more
clustered than the late-forming ones.   We examine the clustering of the
galaxies in these halos later on in \S~\ref{Sec:clustering}.

\begin{figure}
\includegraphics[width=0.48\textwidth]{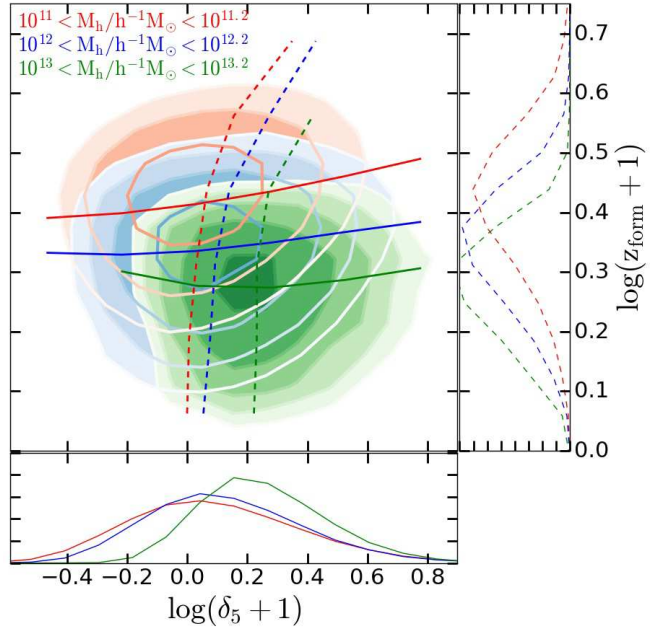}
\caption[]{\label{Fig:Cont}
Joint distribution of large-scale environment ($\rm \delta_5$) and
formation redshift ($\rm z_{\rm form}$) for present-day halos in the 
Millennium simulation, for three narrow ranges of halo mass. The 
red, blue and green contours represent halos with low, intermediate and high 
masses, respectively, as labelled in the top part of the figure. The
different contour levels correspond to 1, 2 and 3 $\sigma$ of the distribution.
The marginalized distributions of each property separately are shown as well,
for each halo mass bin.
The (roughly horizontal) solid lines represent the median values of the 
formation redshift as a function of environment. The (roughly perpendicular) 
dashed lines are the median values of environment at each formation redshift.
}
\end{figure}

To further examine the correlation between formation redshift and large-scale 
environment, we plot in Fig.~\ref{Fig:Cont} the joint distribution of the two
properties. We do this separately for three narrow ranges of halo mass, as 
labelled, since the two properties by themselves also correlate with halo 
mass, which is apparent from the individually marginalized distributions also 
shown. These demonstrate the known trends that more massive halos reside in 
denser environments and are formed later than less massive halos.
The 2D distribution appears very broad with no obvious strong trend. To
quantify that we also plot the medians of one property as a function of the
other: the solid lines are the median of formation redshift for fixed 
density and the dashed lines the median of environment for a given
formation redshift. The fact that the solid lines are roughly horizontal
and the dashed lines nearly perpendicular (or that the two sets of medians 
are almost perpendicular to each other) over most of the range reflects
their lack of correlation on one another.  This is perhaps somewhat surprising
given the measurements of assembly bias (e.g., \citealt{gao05,gao07}) showing
that early-formed halos are more clustered than late-forming ones, and as
such expected to reside in dense environments. Only such a weak dependence 
is apparent, at the high density and high formation redshift end, where the two
sets of lines slightly curve toward each other.    

\vspace{0.1cm}
\subsection{The HOD as a function of halo age and environment}
\label{SubSec:HOD}

It is of fundamental importance and interest to investigate how the halo 
occupation functions themselves vary as a function of each of these 
properties. For the galaxy sets we use fixed number-density samples drawn 
from the SAM catalogs when ranked by stellar mass.  
We have examined a range of different number density samples and present
the results for three representative cases with number densities of
$3.16 \times 10^{-2} \hmpcc$, $ 10^{-2} \hmpcc$ and 
$3.16 \times 10^{-3} \hmpcc$.
The corresponding minimum stellar mass thresholds for each of these are 
provided in Table~\ref{table}.  Naturally, the stellar masses increase with
decreasing number density.  Differences between the stellar mass values
of G11 and L12 are expected, given the differences in galaxy formation
prescriptions and corresponding stellar mass functions. 

\begin{table}
\centering
\caption{\label{table}
Stellar mass thresholds (in units of $h^{-1}M_{\odot}$)
for the three main number-density samples (in units of $\hmpcc$)
presented in this work, for the G11 and L12 models.} 
\hspace{-0.2cm}
\begin{tabular}{c c c c}
\hline
\hline
  & $3.16 \times 10^{-3}$ & $1 \times 10^{-2} $ & $3.16 \times 10^{-2} $ \\ 
\hline
G11    & $3.88 \times 10^{10}$ & $1.42  \times 10^{10}$ & $1.85 \times 10^{9}$\\
L12    & $2.92 \times 10^{10}$ &$6.50 \times 10^{9}$ & $9.39 \times 10^{8}$\\
\hline
\end{tabular}
\end{table}

\begin{figure}
\includegraphics[width=0.48\textwidth]{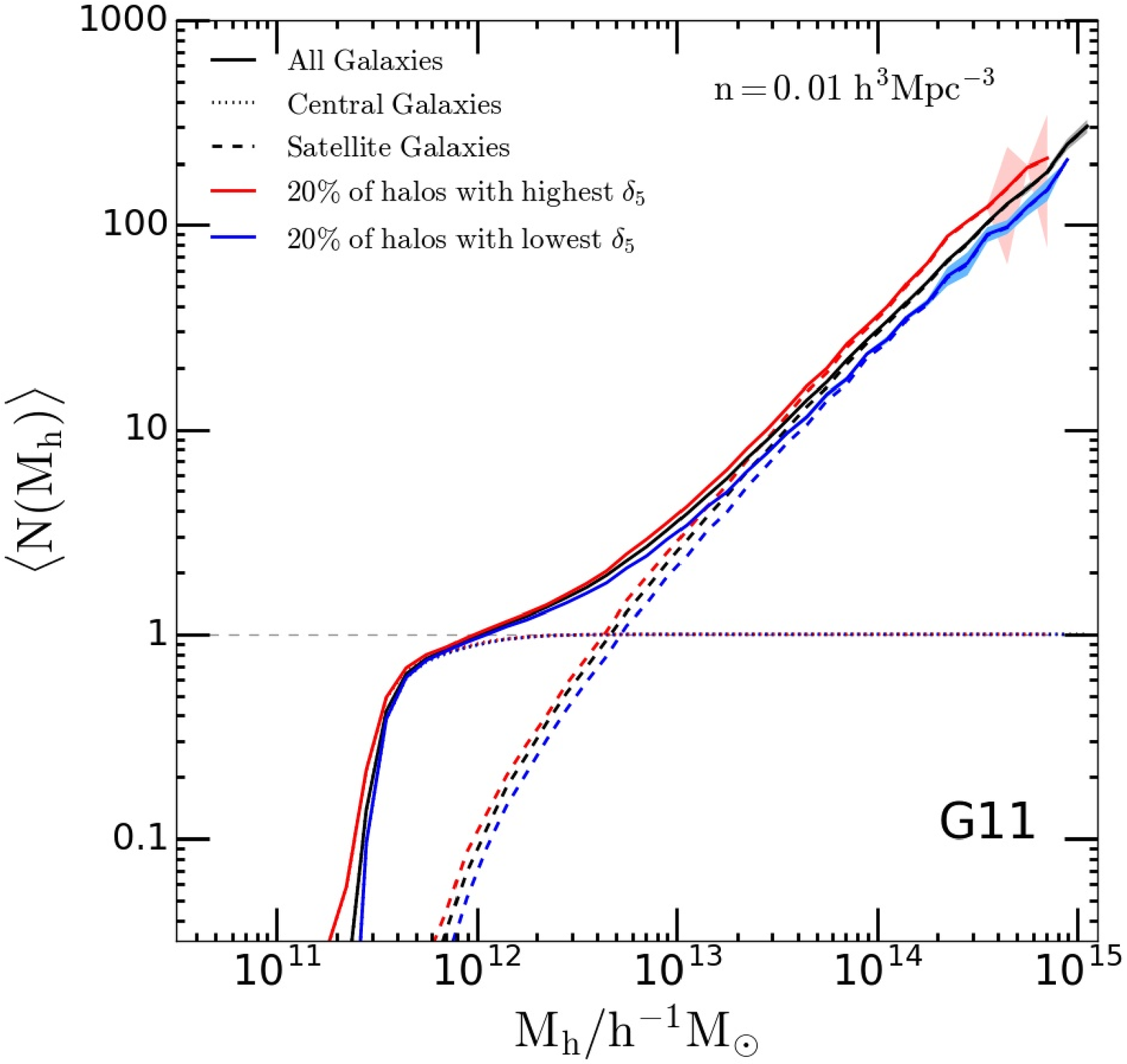}
\includegraphics[width=0.48\textwidth]{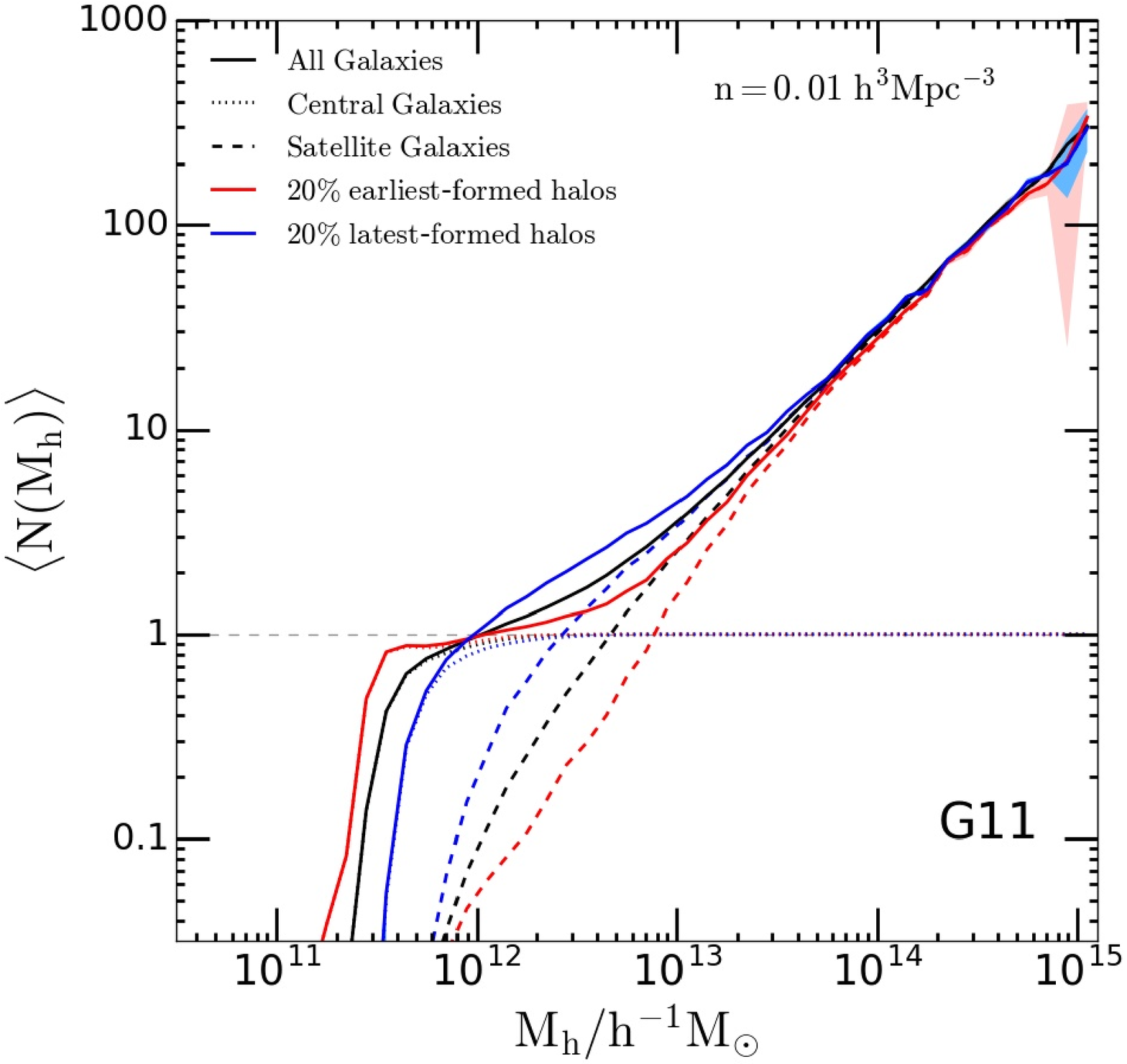}
\caption{\label{Fig:HOD_Err}
(Top panel) The halo occupation functions for a galaxy sample with number 
density $ 10^{-2} \hmpcc$, for the G11 model. The solid black line shows the 
HOD of all galaxies in the sample. The solid red line shows the HOD for the 
galaxies in the 20\% of halos in the densest environments, while the solid 
blue line presents the HOD for the galaxies in the 20\% of halos in the least 
dense environments. The red and blue shaded regions (apparent only at the 
high-mass end) represent jackknife errors calculated using 10 subsamples. In 
all cases, dotted lines show separately the central galaxy occupation 
contribution and  dashed lines represent the satellite galaxies occupation.
(Bottom panel) Same as in the top panel, but here for halo samples selected 
by their formation time instead of their environment. The occupation for 
galaxies in the 20\% earliest-formed halos is shown in red, and for the
20\% latest-formed halos is shown in blue.}
\end{figure}

Fig.~\ref{Fig:HOD_Err} shows how the halo occupation functions vary with 
environment and halo age for a galaxy sample from the G11 SAM model 
with a number density of $10^{-2} \hmpcc$. The top panel shows the HODs
for the full galaxy sample (black) as well as for the subsets of galaxies
that reside in the 20\% of halos in the densest environments (red) and
20\% of halos in the least dense environments (blue). We remind the
reader that the division to 20\% most/least dense regions is done for each
bin of halo mass, so that the different samples equally probe the full
halo mass range. Also, we note that, by construction, these samples have 
equal numbers of halos but not equal number of galaxies. 

We find distinct differences in the HODs for both the central and satellite
occupation functions. For the central occupation, the differences are 
noticeable at the ``knee'' of the occupation function and below. We find that
in the densest environments, central galaxies are more likely to reside also
in lower-mass halos, and the trend reverses in underdense regions. Stated in
a slightly different way, in the regime where the halo occupation rises from
0 to 1, halos are more likely to host central galaxies if they reside in 
dense environments. This may be related to preferential early formation 
of halos in dense regions, though as we saw above the correlation is rather 
loose. We discuss below further insight into  the resulting trends for 
central galaxies (see \S~\ref{Sec:SMHM}).  

The satellite occupation function in the G11 model also exhibits
a dependence on large-scale environment. The satellite occupation function
in dense environments exhibits a slight shift toward larger numbers, so
that halos in dense environments are more likely to have more satellites
on average.  This behavior is perhaps naturally expected, due to the 
increased interactions and halo mergers in dense environments.

The bottom panel of Fig.~\ref{Fig:HOD_Err} shows how the occupation function
varies with halo age for the same G11 galaxy sample with number density 
$10^{-2} \hmpcc$. In the case of halo age, there are much larger effects on 
the occupation functions than we saw with environment. For the central 
occupation, we find a clear trend of early forming (old) halos being more 
likely to host galaxies also at lower masses than late
forming (young) halos.  This likely arises from the fact that the early formed
halos have more time for stars to assemble and for the galaxy to form. The 
sense of the trend is the same as that for the environmental dependence but is
a much more stronger one, with the ``shoulder'' of the occupation function
extending significantly toward lower masses with larger age.

We find a strong reverse effect for the satellites occupation at the
low mass end:  early-forming halos have significantly fewer satellites than
late-forming halos.  This trend is pronounced at low occupation numbers of
$\langle N(M_{\rm h})\rangle<10$ and becomes negligible at higher occupation
numbers.  This is probably due to the fact that in the early-forming halos 
there is simply more time for the satellites to merge with the central galaxy,
which will be a more dominant process at the low halo mass / low occupation
regime.   This trend is similar to the predicted dependence of subhalo
occupation on halo formation time \citep{bosch05,zentner05,giocoli10,jiang16},
indicating that baryonic physics does not play an important role in the
variation of the satellites occupancy.

These differences in the halo occupation functions, for both age and
environment, are significant. We
estimate the uncertainties on the HOD calculations using jackknife 
resampling, dividing the full simulation volume into 10 slices. Incidentally, 
when separating the different subregions, if the center of a given halo is 
in a certain subvolume we include with it all galaxies in that halo, 
regardless of where the physical boundary between the subvolumes lie.  The 
resulting errors are shown as shaded regions in the figure, and are in fact 
negligible over most of the range and only become significant at the 
high-mass range where the number of halos is small.

The HOD dependences on age and environment are different in magnitude 
(for centrals) and sense (for satellites). The strength of the trends with 
age versus environment perhaps indicates that formation time is the 
more fundamental property related to assembly bias. Varying the Gaussian
smoothing length used to define the environment impacts slightly
the size of the deviations, with the differences becoming a bit more 
pronounced for small smoothing lengths, as expected. However, we choose
to stick with our $5 \hmpc$ Gaussian smoothing so as to robustly infer
the large-scale environment.  

We describe and model these differences in terms of the HOD parameters in
\S~\ref{SubSec:param}.   The dependence on environment we find for the
central occupation is very similar to that measured by \citet{mcewen16}.
However, they do not find any noticeable difference for the satellite 
occupation. Our results differ from those of \citet{mehta14} who find no
significant dependence of the HOD on environment. The level of occupancy
variation that is present appears to depend on the specifics of the galaxy
formation model utilized.

\begin{figure}
\includegraphics[width=0.48\textwidth]{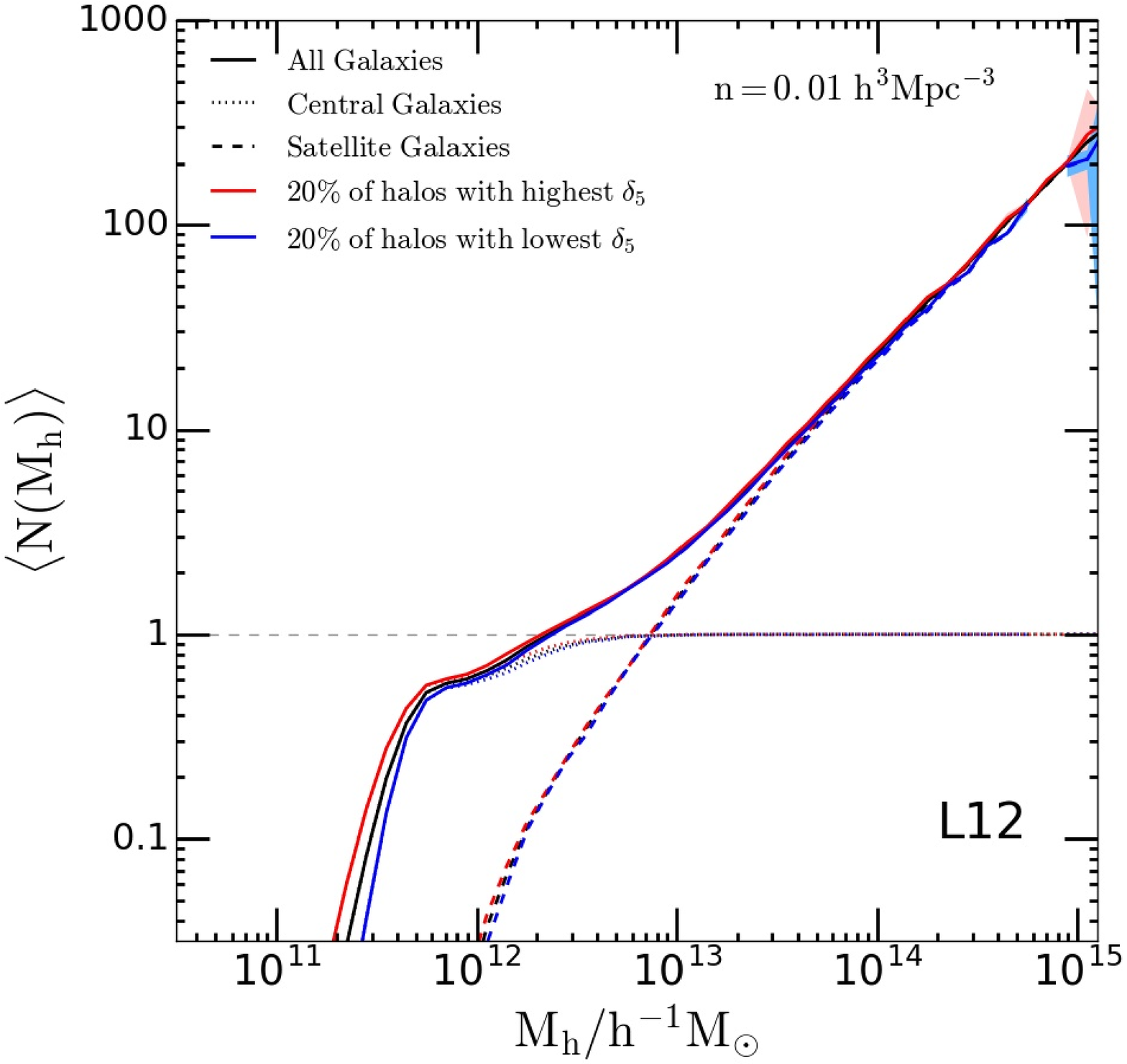}
\includegraphics[width=0.48\textwidth]{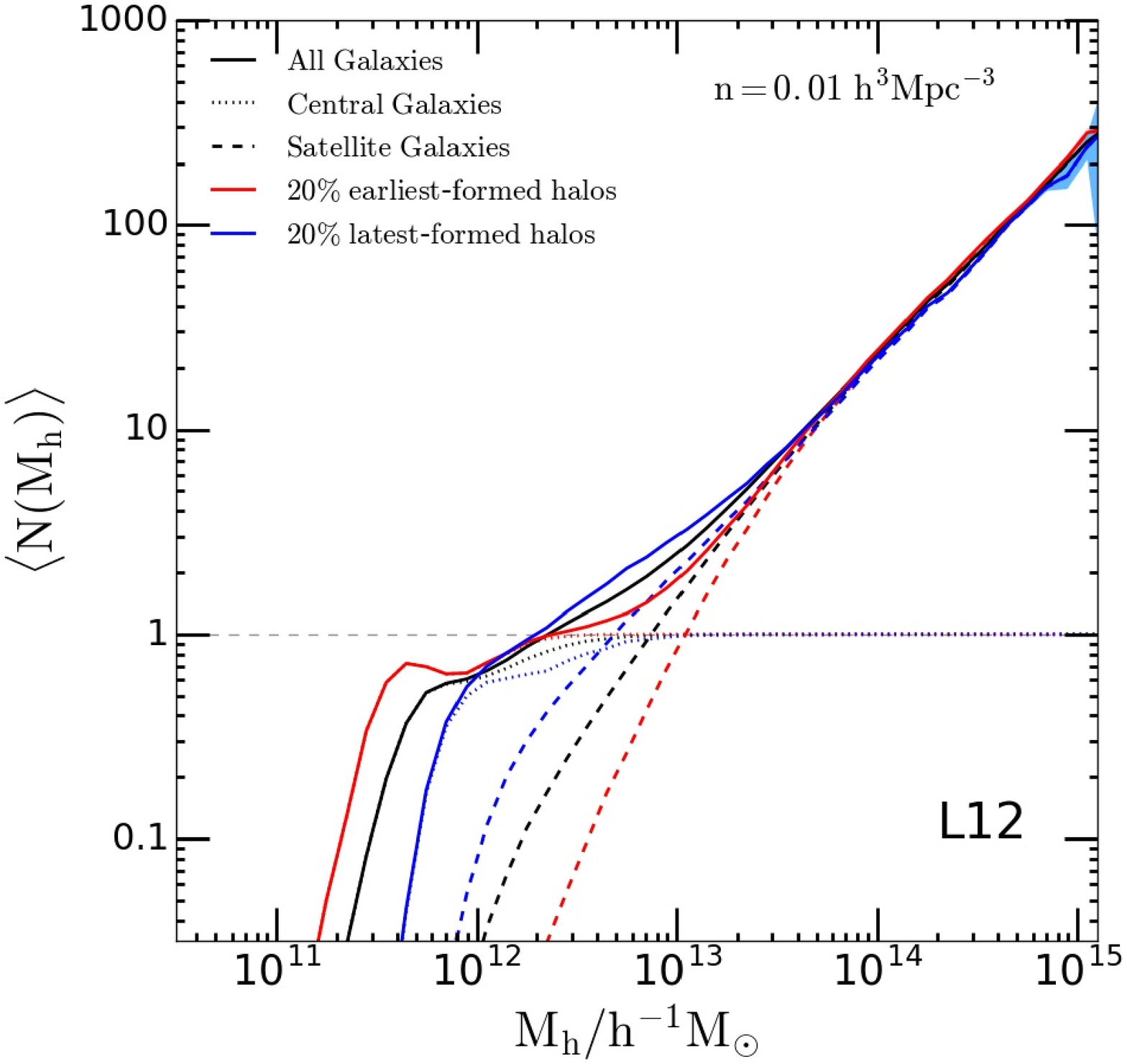}
\caption{\label{Fig:HOD_L12}
The same as Fig.~\ref{Fig:HOD_Err} but for the L12 model.}
\end{figure}

\begin{figure*}
\includegraphics[width=0.48\textwidth]{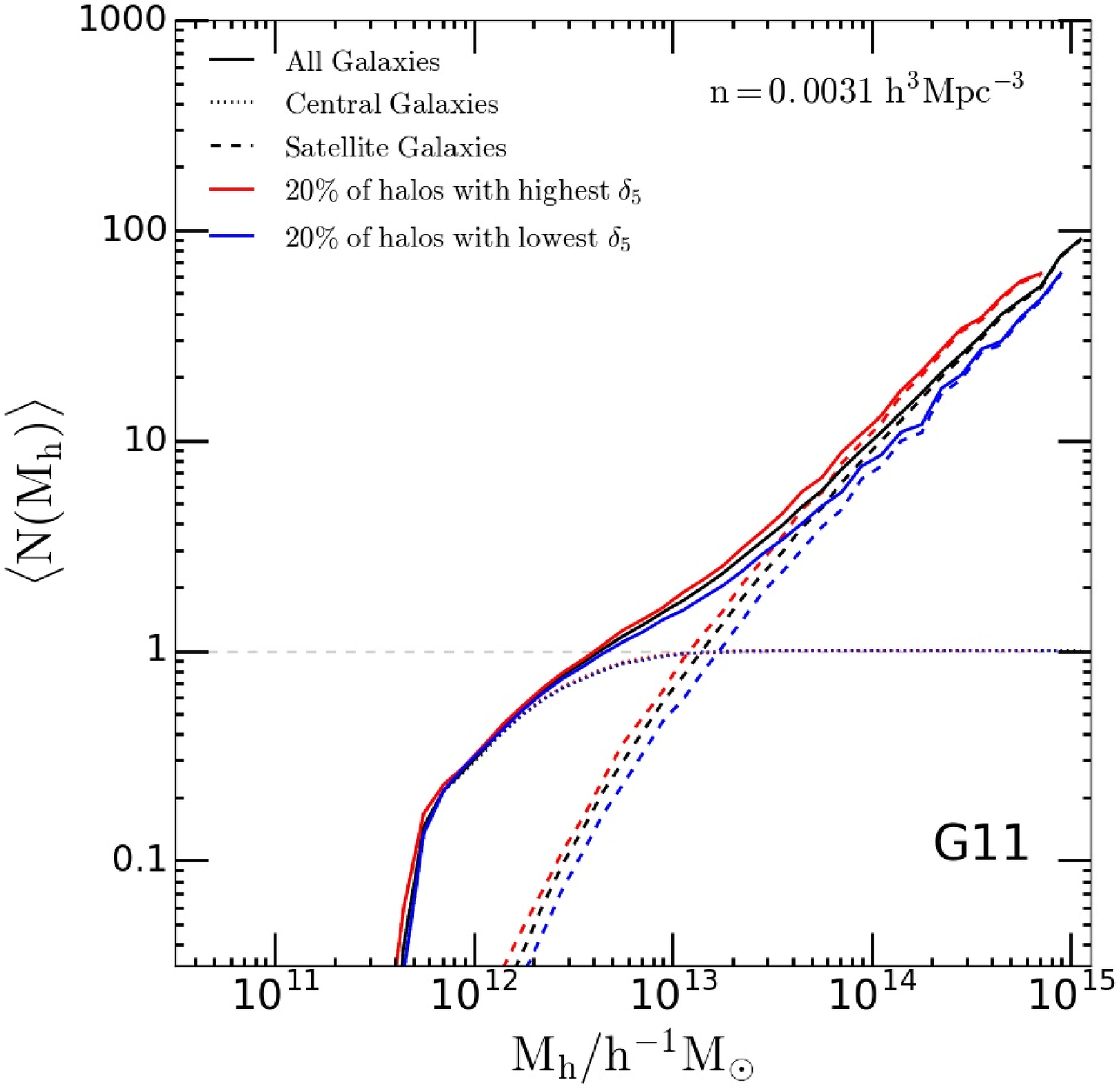}
\includegraphics[width=0.48\textwidth]{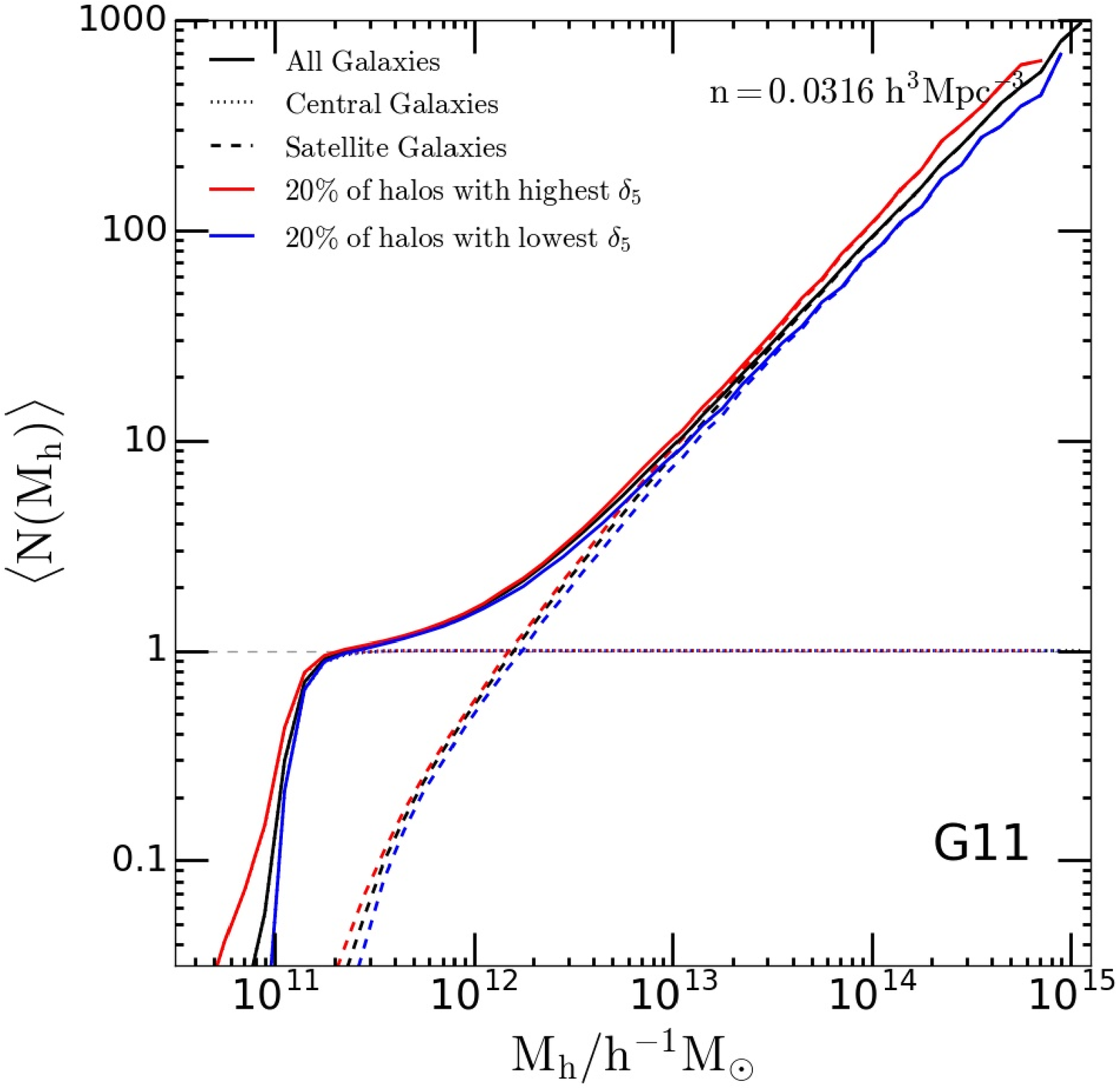}
\includegraphics[width=0.48\textwidth]{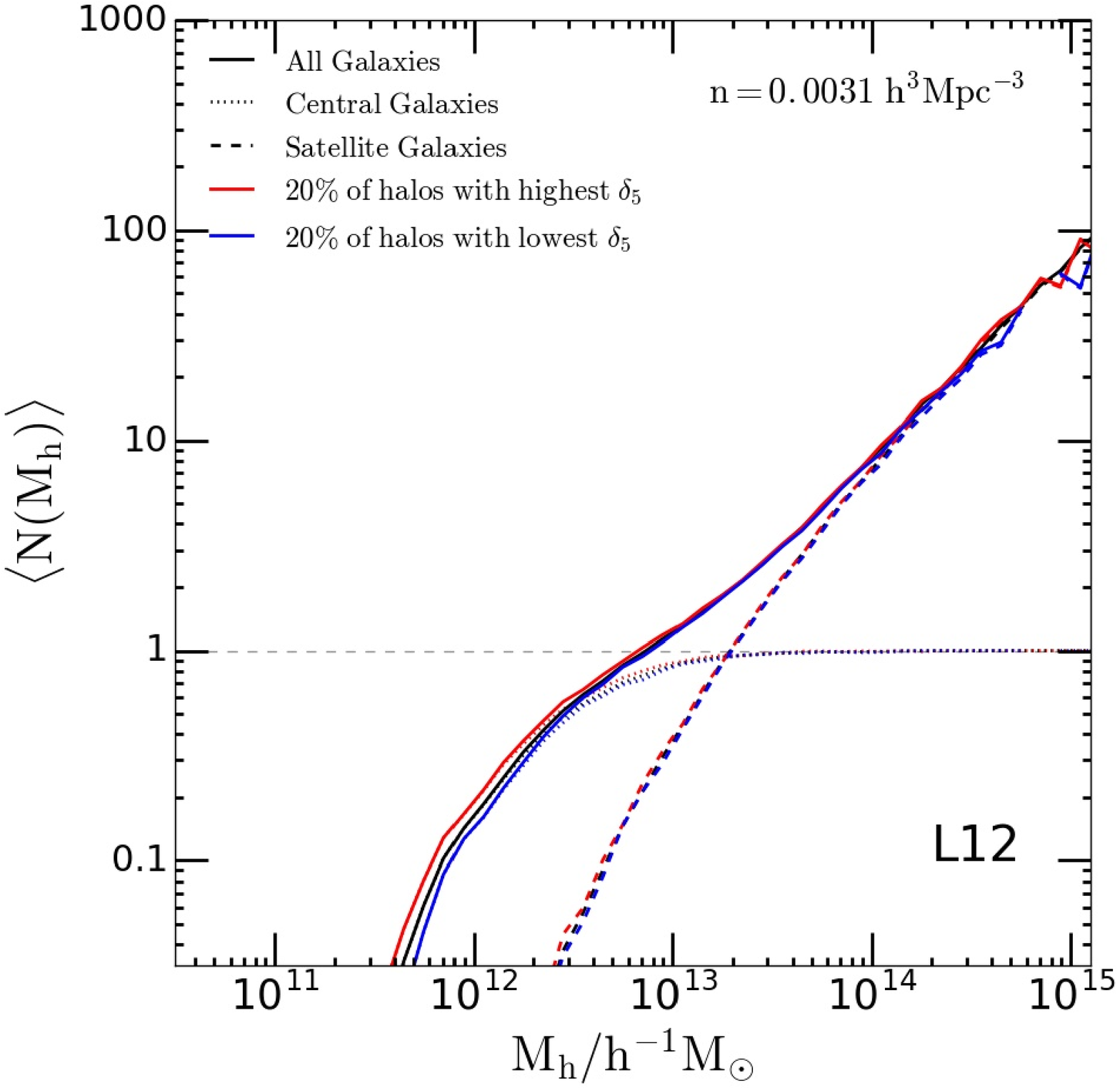}
\hspace{0.6cm}
\includegraphics[width=0.48\textwidth]{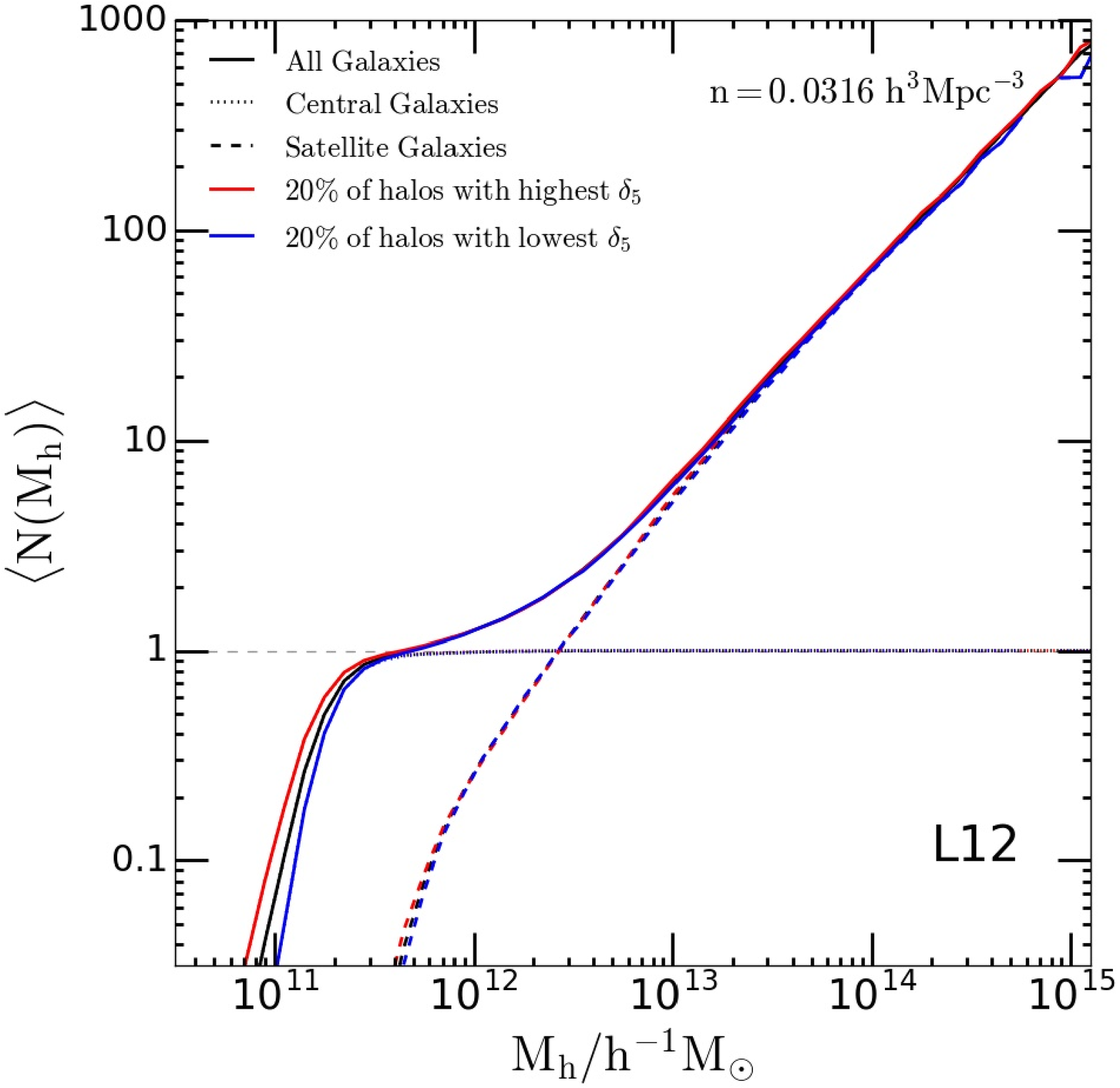}
\caption{\label{Fig:HOD_envir}
The dependence of the halo occupation functions on large-scale environment for
different number densities than that shown in Fig.~\ref{Fig:HOD_Err} and 
Fig.~\ref{Fig:HOD_L12}, $3.16 \times 10^{-3} \hmpcc$ on the left-hand
side  and $3.16 \times 10^{-2} \hmpcc$ on the right. The top panels
are for the G11 model and the bottom ones are for L12.} 
\end{figure*}

\vspace{0.1cm}
\subsection{The HOD for different models and samples}
\label{SubSec:otherHODs}

\begin{figure*}
\includegraphics[width=0.48\textwidth]{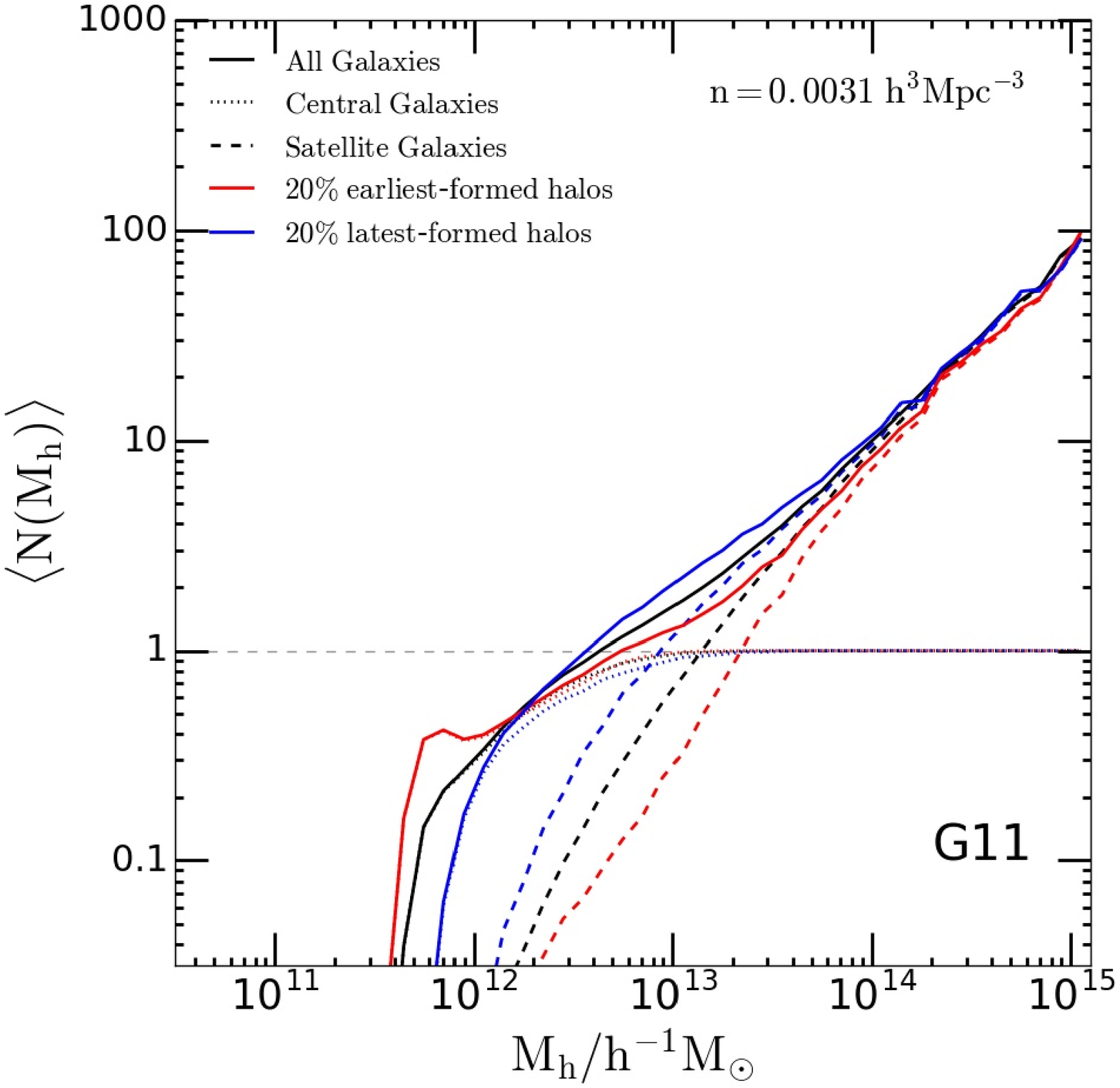}
\includegraphics[width=0.48\textwidth]{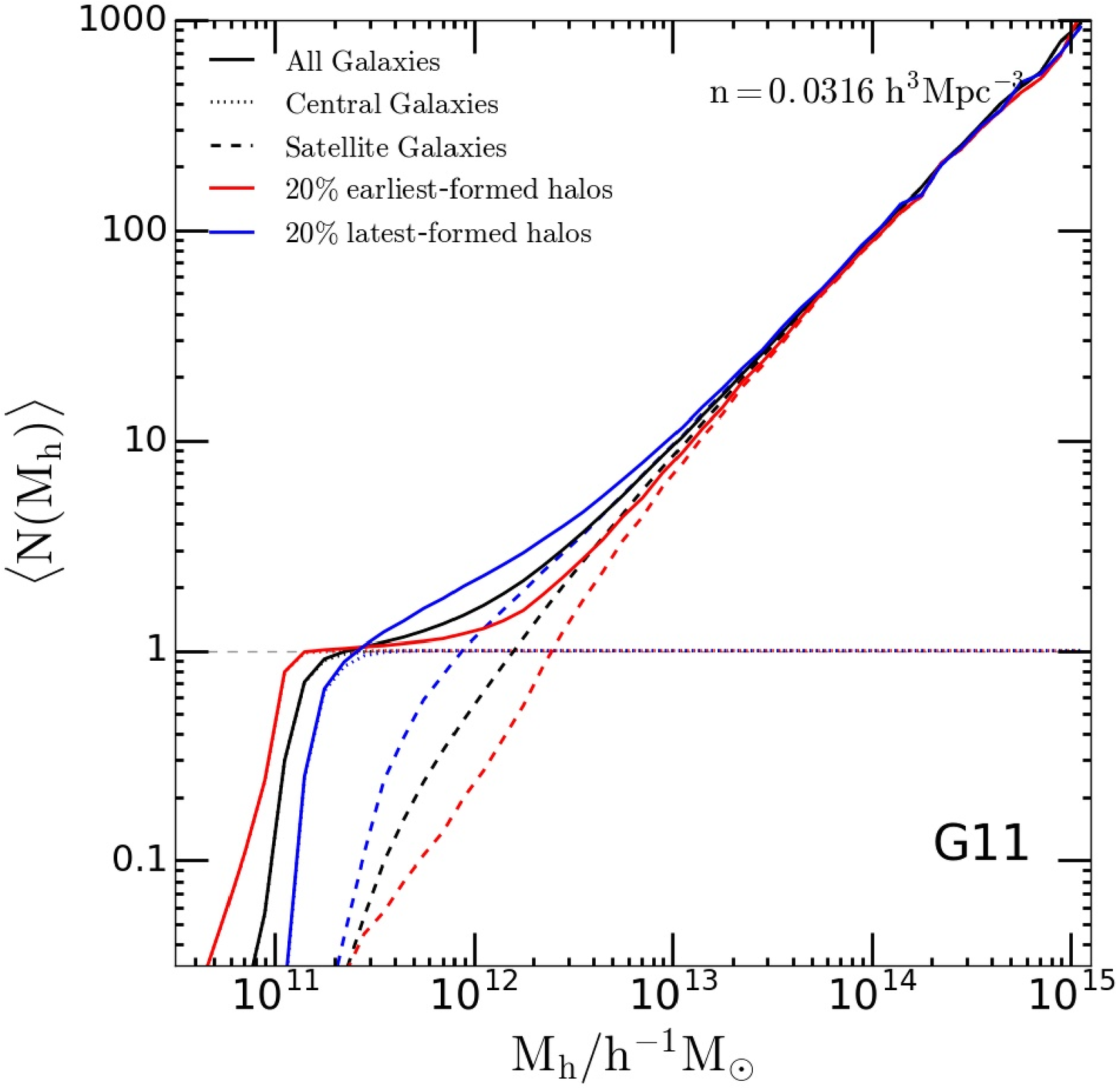}
\includegraphics[width=0.48\textwidth]{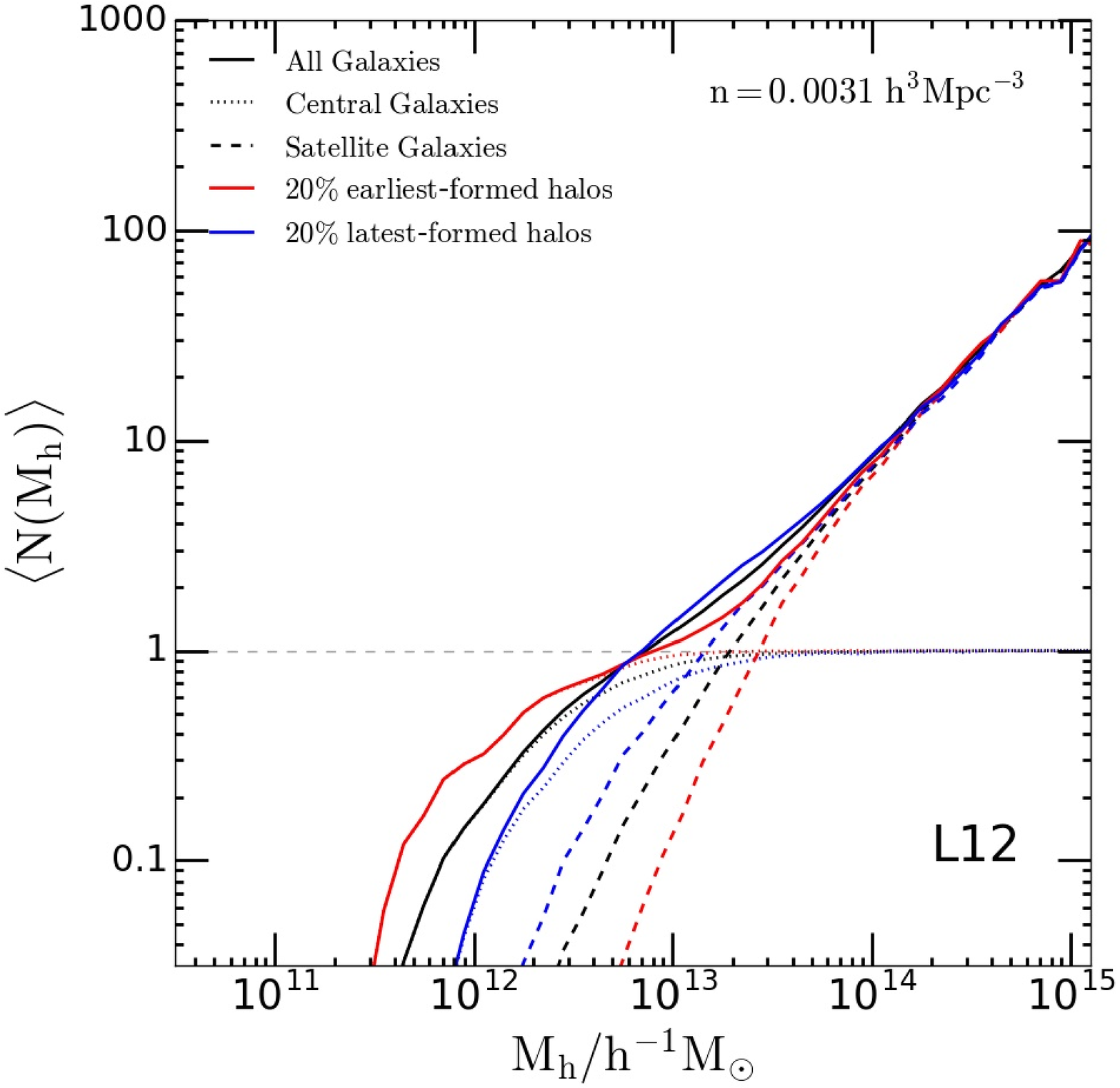}
\hspace{0.6cm}
\includegraphics[width=0.48\textwidth]{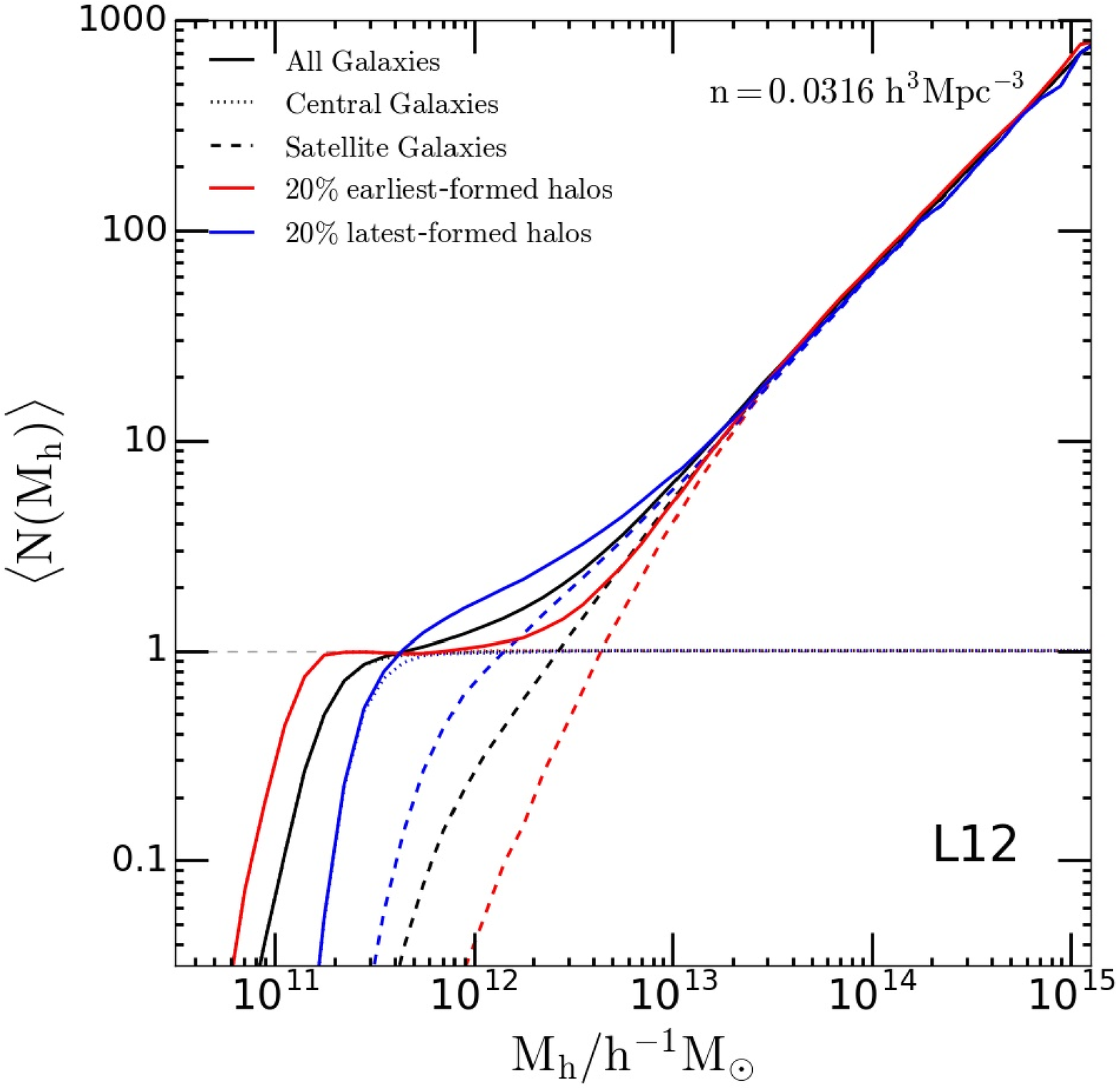}
\caption{\label{Fig:HOD_age}
The same as Fig.~\ref{Fig:HOD_envir}, but for galaxy samples selected using 
halo formation time, instead of environment.}
\end{figure*}

To further investigate the dependence on the galaxy formation model we repeat 
the analysis in \S~\ref{SubSec:HOD} using the independently derived L12 Durham 
model. Figure~\ref{Fig:HOD_L12} shows the HOD dependence on environment and 
halo age for a galaxy sample with number density of  $10^{-2} \hmpcc$ from the
L12 SAM.  The environmental dependence for L12 shows a similar, but more 
subtle, trend for the central occupation, while the trend for the satellite
occupation disappears.  
The difference in the satellite occupations between L12 and G12 could arise 
due to the different treatment of satellites in the two models 
(\S~\ref{SubSec:SAM}). As the satellite destruction processes are more 
immediate in L12, perhaps there is less time for the environmental effects 
to impact the occupation in that case.

The HOD dependence on halo formation time for L12 and G11 is very similar,
with L12 as well showing the strong trends for both the central and satellite
occupation.  The tendency of centrals to shift toward occupying lower
mass halos is slightly stronger for L12. We note the distinct change
of shape of the central occupation, giving rise to a non-monotonic occupation
for the galaxies in early-forming halos. This is likely to be related to the 
form of AGN feedback in the Durham models, as discussed in \citet{nuala17}.

We also examine the dependence of the different trends with stellar mass
of the galaxies, by varying the number density of the samples. As the samples
are ranked by stellar mass, larger number densities include smaller stellar
masses, while small number densities are limited to more massive galaxies.
We present our results for two additional number densities than the one 
previously used (one smaller and one larger) in Figures~\ref{Fig:HOD_envir}
and \ref{Fig:HOD_age}, for environment and age, respectively. In both 
cases, the HODs change globally as expected, shifting overall towards lower 
halo masses with increasing number density (decreasing stellar mass).

The specific signatures of the environmental dependence of the HOD change as 
well with number density. For G11 (top panels of Fig.~\ref{Fig:HOD_envir}), 
the differences in the central occupations 
increase with number density. This is in accordance with the findings 
of \citet{croton07} that galaxy assembly bias is stronger for fainter (less 
massive) galaxies.  For the lowest number density shown, corresponding to 
galaxies with stellar masses larger than $3.88 \times 10^{10} h^{-1}M_{\odot}$ 
(Table~\ref{table}), the differences between the central occupations are
barely noticeable. In contrast, the G11 satellite occupation differences 
decrease slightly with number density.  These opposing changes with
number density suggest that the environment dependence of the central and
satellite occupations  have different origins.
We find a similar change with number density of the central occupation 
environment dependence for L12 (bottom panels of Fig.~\ref{Fig:HOD_envir}), 
while the satellite occupancy variation remains effectively undetected.

The halo age signatures for the different number densities 
(Figs.~\ref{Fig:HOD_Err}, \ref{Fig:HOD_L12} and \ref{Fig:HOD_age}) are quite
robust and do not exhibit any clear dependence on the number density for 
either model, again indicating that these may be of different physical nature. 
The non-monotonic occupation behavior for the early-forming halos 
\citep{nuala17} is also apparent in the smallest number density case for the 
G11 model.

\vspace{0.1cm}
\subsection{Extending the HOD parametrization}
\label{SubSec:param}

\begin{figure*}[bt]
\includegraphics[width=1\textwidth]{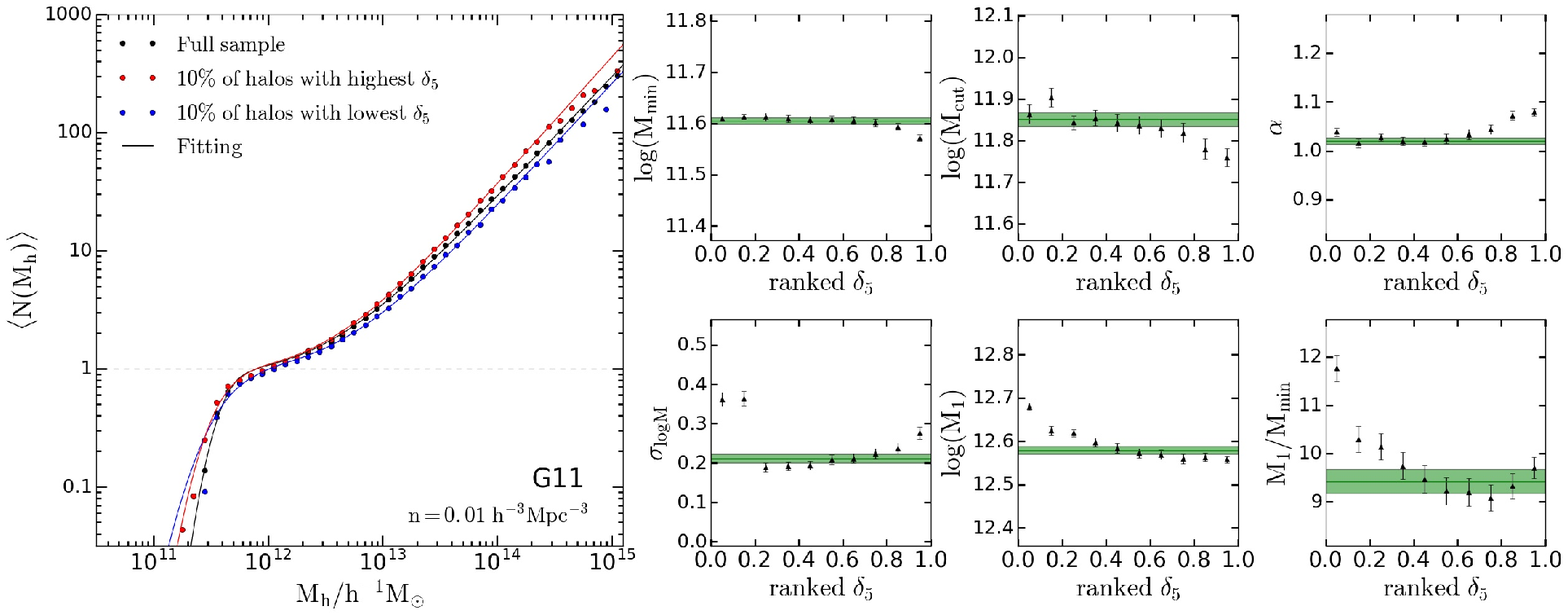}
\caption{\label{Fig:HODfit_Env}
(Left) The HOD of the G11 SAM for a number density of $3.16 \times 10^{-2} 
\hmpcc$. Dots represent the HOD calculated in the simulation: the black ones
show the HOD for all galaxies; the red ones the HOD for the $10\%$ of 
halos in the densest environments; and the blue ones show the HOD for the 
$10\%$ of halos in the least dense environments. The solid lines, in 
corresponding colors, show the 5-parameter best-fit models for these.
(Right) The values of the best-fitting parameters of the HODs as a function
of the environment percentile for $M_{\rm min}$ (top left), 
$\sigma_{\rm logM}$ (bottom left), $M_{\rm cut}$ (top middle), $M_{\rm 1}$ (bottom
middle), $\alpha$ (top right) and $M_{\rm 1}/M_{\rm min}$ (bottom right).  Each 
dot in these plots represents a 
different subsample selected by its large-scale environment, each with 
$10\%$ of the full halo population, with the environment density increasing
from left to right. The left-most dots and right-most dots in these panels 
represent the parameter values of the models plotted in blue and red, 
respectively, in the left-hand side HOD panel.  The errorbars reflect the
$1\sigma$ uncertainty on the parameters.
The green horizontal lines with shaded regions in the parameters panels are
the values fitted for the full sample and their uncertainty.
} 
\end{figure*}

\begin{figure*}
\includegraphics[width=1\textwidth]{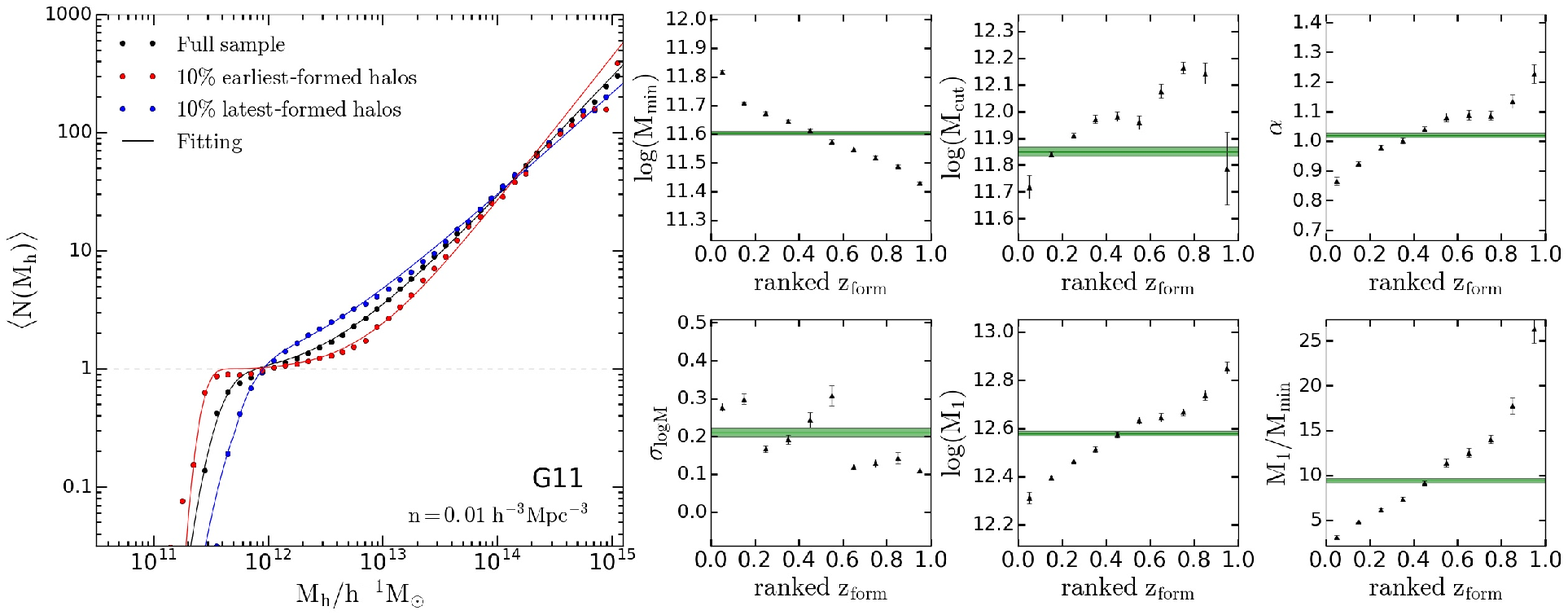}
\caption{\label{Fig:HODfit_Age}
Same as Fig.~\ref{Fig:HODfit_Env}, but for subsamples selected by  
halo formation time instead of large-scale environment. In the panels for
the individual parameters, the halo formation redshift (age) increases 
going from left to right. Please note that for the parameter values on
the right, the y-axis ranges are different than those of
Fig.~\ref{Fig:HODfit_Env}.}
\end{figure*}

It is customary to parametrize the shape of the HOD using a 5-parameter model
which captures the main features of the halo occupation function, as predicted
by SAMs and hydrodynamic simulations \citep{zheng05}. This model is commonly
used when interpreting galaxy clustering measurements to infer the galaxy--halo
connection (e.g., \citealt{zheng07,zehavi11}).  Here, we characterize the
HOD dependences on age and environment in terms of the 5 parameters, as a 
first step toward incorporating these variations into the HOD model.

The halo occupation function is usually modeled separately for 
central galaxies and satellites. The occupation function for
centrals is a softened step-like function with the following form:
\begin{equation}
 \langle N_{\rm cen}(M_{\rm h})\rangle = \frac{1}{2}\left[ 1 + {\rm erf} \left( \frac{\log M_{\rm h} - \log M_{\rm min}}{\sigma_{\log M}}  \right) \right],
\label{Eq:Cen_HOD}
\end{equation}
where $ {\rm erf}(x)$ is the error function,
$ {\rm erf}(x) = \frac{2}{\sqrt{\pi}} \int_{0}^{x} e^{-t^2} {\rm d}t. $
$M_{\rm min}$ characterizes the minimum halo mass for hosting a central galaxy
above the specified threshold. 
In the form adopted here, it is the halo mass for which half of the halos 
are occupied.  $\sigma_{\log M}$ indicates the width of the transition from 
zero to one galaxy per halo and reflects the scatter between stellar mass 
and halo mass.

For satellite galaxies, the occupation function is modeled as:
\begin{equation}
 \langle N_{\rm sat}(M_{\rm h})\rangle = \left( \frac{M_{\rm h}-M_{\rm cut}}{M^*_1}\right)^\alpha,
\label{Eq:Sat_HOD}
\end{equation}
for $M_{\rm h}>M_{\rm cut}$,
representing a power-law occupation function with a smooth cutoff at the 
low-mass end. Here $\alpha$ is the slope of the power-law, with typical
values close to one,  
$M_{\rm cut}$ is the satellite cutoff mass scale (i.e., the minimum mass 
of halos hosting satellites), and $M^*_1$ is the normalization.
Often, instead of the latter, a related parameter is used, $M_{1}$, which
is the mass of halos that host one satellite galaxy on average 
($M_{1} = M^*_1 + M_{\rm cut}$).
The total occupation function is then specified by these 5 parameters and
given by the sum of the two terms:
\begin{equation}
 \langle N_{\rm gal}(M_{\rm h})\rangle =  \langle N_{\rm cen}(M_{\rm h})\rangle +  \langle N_{\rm sat}(M_{\rm h})\rangle.
\end{equation}

Figure~\ref{Fig:HODfit_Env} shows how these 5 parameters vary with environment.
The left-hand side presents the HOD of the G11 SAM for 
$n=3.16 \times 10^{-2} h^{3} Mpc^{-3}$ for the full sample, and the $10\%$ of
halos in the most dense regions and the $10\%$ of halos in the least dense
regions. The dots represent the directly-measured HODs and the lines are
the best-fit 5-parameter models to them. The right-hand side examines how 
each of the parameters varies with environment in $10\%$ bins of halo 
environment.  The fits are done assuming equal weight to all measurements and 
using only those with  $\langle N(M_{\rm h})\rangle > 0.1$.  The errorbars on
the parameters are obtained by requiring $\chi^2/{\rm dof}=1$, as in 
\citet{Contreras17}.  

For this G11 sample, we see that the changes to the parameters when varying
the environment are subtle, but all are affected.
The changes in the central occupation with density 
are in fact quite small, with $M_{\rm min}$ decreasing and $\sigma_{\rm logM}$ 
increasing slightly with density.  The variations in best-fitting 
parameters are influenced by the limited flexibility in the assumed shape of 
the HOD.
The changes in the satellite occupation with increasing density act to
gradually decrease $M_{\rm cut}$ and $M_{\rm 1}$ and increase the slope 
$\alpha$, over at least part of the density range. 
We note that we find more intricate changes to the HOD parameters than those
modeled in \citet{mcewen16}, since that work saw differences only in the
centrals occupation function and not the satellites one. The resulting
variation in the $M_{\rm 1}/M_{\rm min}$ ratio (bottom-right panel) is a 
noticeable decrease with increasing density over most of the range, but 
then a turnover and a slight increase for halos in the densest regions.

Fig~\ref{Fig:HODfit_Age} examines the change in the parameters, but now
with halo formation time. The change in parameters in this case is more 
distinct and significant, since the dependence of the HOD on halo age is 
stronger than on environment. $M_{\rm min}$ monotonically decreases with 
increasing formation redshift (earlier formation). $\sigma_{\rm logM}$ varies 
with halo age but does not show a clear trend.
The satellite occupation changes in the
opposite sense, with all three parameters $M_{\rm cut}$, $M_{\rm 1}$ and 
$\alpha$ increasing significantly with larger formation redshift. 
(The change in the slope $\alpha$ again may be somewhat
affected by the limitations of the assumed HOD shape.)

The combined effect on the $M_{\rm 1}/M_{\rm min}$ ratio is a dramatic 
increase with formation redshift, of about a factor six over the full range!
This change is much stronger than the variation of this ratio
with either number density or redshift, as explored by \citet{Contreras17}, 
about twice as large as the variation with number density and close to four 
times larger than the evolution in the ratio from redshift 3 to 0.
This significant change, however, is easily understood from the predicted 
occupancy variation (e.g., bottom part of Fig.~\ref{Fig:HOD_Err}). 
For earlier-forming halos, $M_{\rm 1}$ shifts toward larger halo masses 
while $M_{\rm min}$ shifts toward smaller halo masses, resulting in a 
substantial increase in their ratio. Still, it is noteworthy that the 
$M_{\rm 1}/M_{\rm min}$ ratio is such a sensitive indicator of halo age. 

These results can inform theoretical modeling efforts extending the standard
HOD framework. We can envision modeling each of the parameters change with 
halo age as a power-law function with an additional assembly bias parameter 
(similar to our modeling of the evolution of the HOD in 
\citealt{Contreras17}).  Such a model may aid in obtaining constraints on 
assembly bias from observational data, as well as providing a straight-forward
method of incorporating the age-dependence of the HOD into galaxy mock 
catalogs. 

\vspace{0.2cm}
\section{The stellar mass -- halo mass relation}
\label{Sec:SMHM}

To gain a better understanding of the origin of the trends seen in the 
central galaxy occupation function with age and environment, we examine 
the stellar mass -- halo mass (SMHM) relation. As we 
show, it is the dependence of the scatter in this relation on the secondary
parameters that gives rise to the occupancy variation and to galaxy assembly
bias.

\begin{figure}
\hspace{-0.2cm}
\includegraphics[width=0.5\textwidth]{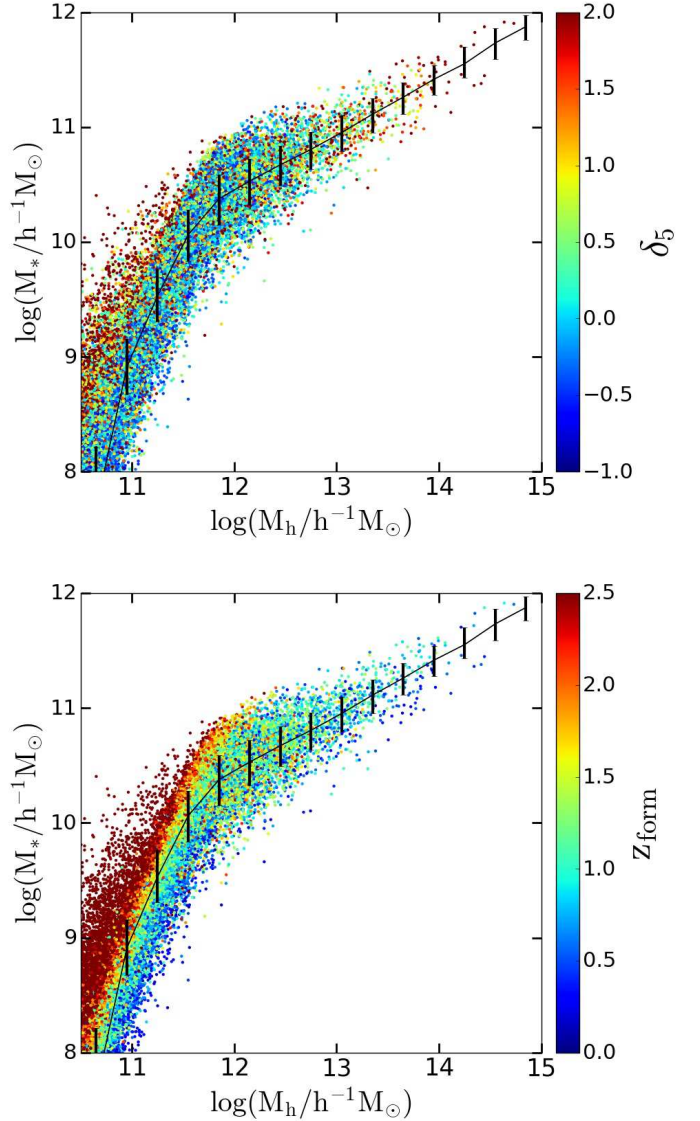}
\caption{\label{Fig:SMHM}
(Top) The stellar mass of central galaxies as a function of host 
halo mass for the G11 SAM and its dependence on environment.
Each dot represents a central galaxy, plotted for a representative
(randomly chosen) $1\%$ of the galaxies. 
Galaxies are color-coded by their $5 \hmpc$ Gaussian smoothed 
density, $\delta_5$, according to the color scale shown on the right. 
The plotting order is also chosen randomly, so as to avoid any overplotting 
issue. The black solid line represents the median of the distribution with the 
errorbars designating the $20\%$--$80\%$ range of the distribution. 
(Bottom) Same as for the top panel, but now color-coded by the formation 
redshift, calculated as the time when the halo reaches half of its final mass. 
For fixed halo mass, more massive central galaxies tend to live in halos that 
formed early or reside in denser environments (with the latter being a weaker 
trend than the former).
}
\end{figure}

Figure~\ref{Fig:SMHM} shows the stellar mass of central galaxies as
a function of halo mass for galaxies in the G11 SAM. We plot $1\%$ of
all central galaxies, for clarity. The stellar mass
increases with halo mass, with the median of the relation (black line)
exhibiting a relatively steep slope up to $M_{\rm h} \sim 10^{12} h^{-1}M_{\odot}$ 
and a shallower increase for more massive halos, when the AGN feedback
becomes important. This was studied in detail in \citet{Mitchell16} for
{\tt GALFORM} (see also \citealt{Contreras15}).  There is significant 
scatter in the relation which decreases at the high-mass end. This scatter 
is expected to be due to stochasticity in both galaxy and halo 
assembly histories and the various physical processes. Thus we may expect 
the scatter to relate to the properties of the host halos. We examine this 
visually by color-coding each galaxy by its large-scale environment (top
panel) and by the formation redshift of its host halo (bottom panel).

As is apparent from Fig.~\ref{Fig:SMHM} the spread around the median SMHM 
relation is not random, but depends on the secondary property. For halos less 
massive than about $10^{12} h^{-1} M_{\odot}$, there is an apparent dependence
on the large-scale environment (top panel), where for fixed halo mass, more 
massive central galaxies tend to reside in denser environments. This trend
appears to have a fairly sharp transition between relatively low and high 
densities, even though there is a large scatter of different environments 
at each location on the SMHM relation (as is evident by the mix of colors).  
The trend does not persist toward larger masses, where there is no variation of 
environment for fixed halo mass (or else it is impossible to see one due to
the large scatter of different densities). 
We find that the central galaxies in the densest environments (in 
absolute terms, not per halo mass bin, i.e., the red/maroon colored ones 
in the top panel) populate two distinct regions in this diagram: they 
predominantly populate the most massive halos, albeit at smaller numbers
according to the halo mass function, and they also comprise the most massive 
centrals in low-mass halos. The former simply stems from the fact that the 
most massive halos tend to reside in dense environments, while the latter 
is related to the occupancy variation we discuss here.

The bottom panel of Fig.~\ref{Fig:SMHM} shows the same SMHM relation, 
but now color-coded by the formation redshift (age) of the halos. The trend 
with halo age for fixed halo mass is particularly striking, with
more massive central galaxies generally residing in halos that formed early.
This dependence on halo age is gradual but very distinct, due to the small
scatter of halo ages at each location in the stellar mass -- halo mass
diagram for halos below 
$\sim 10^{12} h^{-1} M_{\odot}$. The trend persists for all halo masses,
but with a significantly larger scatter of halo ages at the high-mass end, 
as the formation redshifts also progressively become more recent, as expected.
A similar trend with halo formation
time has been measured in  SAMs by \citet{wang13a} and more recently
also for galaxies in the EAGLE hydrodynamical simulation \citep{matthee17}.
It likely arises because central galaxies in early-formed halos have more
time for accretion and star formation and thus end up being more massive.
Once again it appears that halo age is the more fundamental characteristic
here that affects galaxy properties.

The dependence on environment is more complex and harder to interpret.
\citet{jung14} investigate the stellar mass dependence on environment for
fixed halo mass using a different SAM. They find only small differences 
between halos in the densest and least dense environments for low
halo masses, and these differences diminish with increasing halo mass 
(cf. \citealt{tonnesen15}). This suggests that the level of secondary 
correlations present (and by association, galaxy assembly bias) depends 
on the details of the galaxy formation model adopted.
We note also the counter-intuitive fact that, at least according to the study 
of \citet{matthee17},  while some fraction of the scatter in the SMHM 
relation is accounted for by formation time, the large-scale environment 
seems to make a negligible contribution.

The fundamental importance of these dependences of stellar mass on secondary
properties at fixed halo mass is that they provide a direct explanation
for the central galaxy occupancy variation with environment and halo age 
(as shown in, e.g., Fig.~\ref{Fig:HOD_Err}). For fixed halo mass, early-formed 
halos or halos in denser environments host more massive galaxies. Consequently,
any fixed stellar-mass cut (e.g., the $1.42 \times 10^{10} h^{-1}M_{\odot}$
threshold used to define the sample analyzed in Fig.~\ref{Fig:HOD_Err}) 
would include these first.  Thus the central galaxies in early-forming halos 
or dense environments populate relatively lower-mass halos, extending the 
central occupation function in that direction.
And, conversely, late-forming halos or halos in underdense
environments generally host lower-mass galaxies. Therefore, only centrals
hosted by more massive halos will make it into the sample and the central
occupation function in that case will be shifted toward more massive halos. 

The level of scatter in halo age or environment at each location directly
determines the strength of the occupancy variation. The tight correlation
between stellar mass and halo age (for fixed halo mass) results in a large
variation of the HOD,  while the large scatter involved with environment
results in only a moderate change of the HOD in that case.
Furthermore, as noted already, the SMHM trend with environment holds only 
at the low-mass end, while the general trend with halo age persists for all 
halo masses. This explains the change in occupancy variation with number
density, demonstrated in Fig.~\ref{Fig:HOD_envir} and \ref{Fig:HOD_age}. 
For age, the occupancy variation remains at comparable levels for all
number densities, similar to the trend in the SMHM relation.
For environment, the level of occupation variation decreases for smaller
number densities (larger stellar mass thresholds), as these correspond to
larger halo masses where the trend with environment diminishes.

In any case, we are seeing that the correlated nature of the scatter in
the SMHM relation is intimately related to the trends
in the occupation functions.  It is exactly this coupling between halo  
properties (such as large-scale environment and formation time) and galaxy
properties (such as stellar mass or luminosity) that causes the dependence 
of the HOD on halo assembly.
A more extensive study of the connection between the SMHM relation and the 
occupancy variation and galaxy assembly bias will be presented elsewhere
(Zehavi et al., in preparation).

\vspace{0.2cm}
\section{The impact on galaxy clustering} 
\label{Sec:clustering}
To see the impact of the occupancy variation with halo age and environment 
(\S~\ref{Sec:HOD}) on galaxy clustering, we measure and examine the 
correlation functions of galaxies in these samples. The variations in the 
HODs couple with the different clustering properties of the halos to produce 
a signature of assembly bias in the galaxy distribution.

\vspace{0.1cm}
\subsection{The shuffling mechanism}
\label{SubSec:Shuffle}

To measure the effects of assembly bias on the galaxy correlation function, 
we need to create a control sample of galaxies where we explicitly remove
the galaxy assembly bias, and then compare to the clustering of the
original sample. In order to do that we shuffle the full galaxy population 
among halos of similar masses, following the  procedure of \citet{croton07}. 
Specifically, we select halos in 0.2 dex bins of halo mass and randomly 
reassign the central galaxies hosted by these halos among all halos in that 
mass bin, placing them at the same location as the original central galaxy 
in the halo. The satellite galaxies are moved together with their original 
central galaxy, preserving their distribution around it.
The shuffling eliminates a dependence of the galaxy population on any 
inherent properties of their host halos other than mass (since these are 
now randomly assigned).
Practically, what the shuffling does is remove the occupancy variation,
namely the dependence of the HOD on halo properties other than mass.
For these shuffled samples,  the HOD of the full galaxy sample remains 
the same, but the differences between the HODs of different halo populations, 
e.g., split by age or environment, are now eliminated and all share the
same HOD as that of the full sample.

We have verified that the results we present below are insensitive to our
specific choice of the bin size in halo mass. We also note that alternative 
shuffling 
algorithms have been proposed in the literature, where the satellites are 
also shuffled among different halos of the same mass independent of the 
central galaxies (e.g., \citealt{Zu08,zentner14}). This additional satellite 
shuffling is important only when one is specifically concerned with features 
that correlate the properties
of centrals and satellites or the satellites with themselves, such as 
galactic conformity or satellite alignment.  For our purposes, the 
combined central+satellite galaxies shuffling completely suffices to erase the 
signature of the occupancy variation.  We clarify that our shuffling does
impact the small scales (1-halo term) of the correlation function of our
subsamples, as we show below (in contrast to the statement made in some works
that this shuffling preserves the 1-halo term, which only holds when 
considering the {\it full} galaxy sample).

\vspace{0.1cm}
\subsection{The correlation functions}
\label{SubSec:clustering}

\begin{figure}
\includegraphics[width=0.48\textwidth]{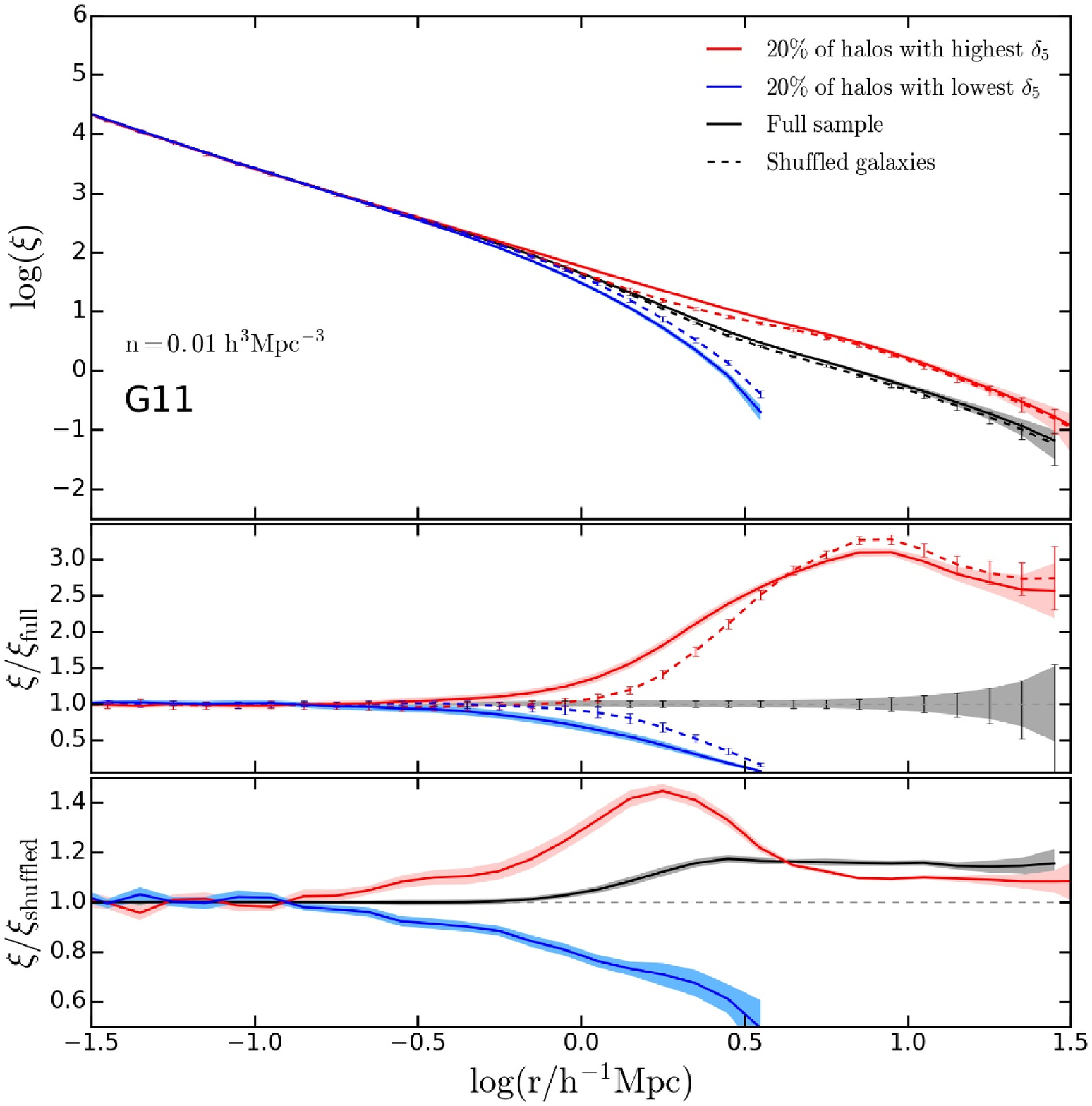}
\includegraphics[width=0.48\textwidth]{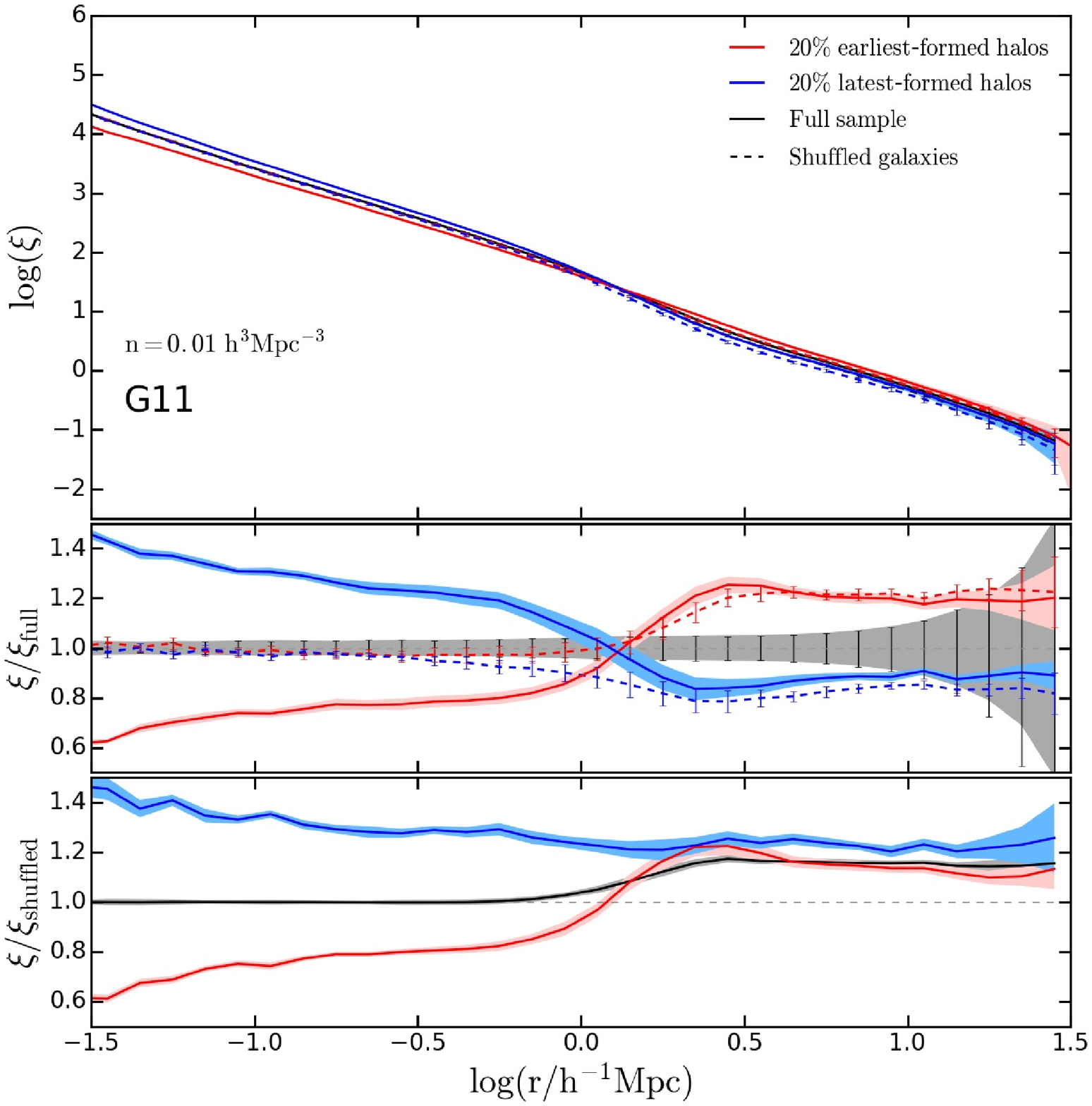}
\caption{\label{Fig:CF_Err}
Correlation functions for the G11 $n=10^{-2} \hmpcc$ sample.
(Top panel) 
The top subpanel shows the auto-correlation function of the full galaxy 
sample (black solid line) and the cross-correlation functions of the
full sample and galaxies in the 20\% of halos in the most and least dense 
regions (solid red and blue lines, respectively). 
Dashed lines are the corresponding correlation functions 
of the shuffled galaxies (see text).  The middle subpanel displays these 
now divided by the full sample auto-correlation function, highlighting 
the different clustering properties in different environments.  This is 
shown for both original and shuffled galaxy samples, e.g., 
the dashed red line is the ratio of the cross-correlation of the shuffled
galaxies in the dense regions and the auto-correlation of the full shuffled
sample.  The bottom subpanel shows, for the three
cases, the ratio between the correlation functions of the original and 
shuffled galaxy samples.   In all subpanels, the shaded regions represent
the errorbars estimated from 10 jackknife realizations.
(Bottom panel) Same as in the top panel, but now for galaxies residing in
halo samples chosen by their formation redshift instead of environment. 
Red (blue) lines correspond to the cross-correlation function 
of the full galaxy sample with the subset of galaxies residing in the 20\% 
earliest- (latest-)formed halos.
}
\end{figure}

Figure~\ref{Fig:CF_Err} show the correlation functions measured for the
galaxy subsamples analyzed in Fig.~\ref{Fig:HOD_Err}, for the
$n=10^{-2} \hmpcc$ sample from the G11 SAM.  We calculate the 
auto-correlation function of the full galaxy sample (solid black lines) and
the cross-correlation function between the full sample and the galaxies 
in the different subsets of 20\% of the halos (red and blue solid lines, as
labelled), showing the environmental dependence on the top and formation time
on the bottom.  The dashed lines in all cases show the results when 
shuffling the galaxy samples, effectively removing the occupancy variation,
as described in \S~\ref{SubSec:Shuffle}. The top subpanels are the correlation
functions themselves, and the middle and bottom subpanels show ratios
derived from these correlation functions, to highlight different features
as described in the figure caption and discussed below.

The shaded regions represent the uncertainties on these measurements 
estimated from jackknife resampling, when dividing the full simulation volume 
into 10 slices along one axis (identical subvolumes to those used to estimate 
the errors in Fig.~\ref{Fig:HOD_Err}).  
The uncertainties on the ratios in the middle and bottom subpanels are the 
jackknife errors on the ratios themselves, which are significantly smaller 
than propagating the individual measurement errors, as can be seen by
comparing the red and blue shaded regions in the middle subpanels with the
grey shaded regions in the middle subpanels (the jackknife measurement errors 
for the full sample). This is expected as the variations of the different 
auto-correlation and cross-correlation functions among the different 
jackknife samples are naturally correlated.
(Note that the y-axis range in the two subpanels is different and the grey 
uncertainty regions plotted in the top and bottom parts of the figure are 
identical.)

We start by examining the top part and top subpanel of Figure~\ref{Fig:CF_Err} 
which illustrates the dependence of clustering on environment.  We find 
distinct differences on large scales between the clustering of the galaxies 
in the most dense environments versus the least dense regions. The galaxies 
in dense environments are significantly more clustered than the full sample, 
while the galaxies in the underdense regions are much less clustered, as 
expected. The cross-correlation functions do not have the same shape on large 
scales as the full auto-correlation function, due to the way these samples 
were defined using the $5 \hmpc$ Gaussian smoothed density fields, 
which effectively carves out different regions of dense and underdense 
environments as is seen in Fig.~\ref{Fig:CW}.  In particular, the 
cross-correlation function for the underdense regions exhibits a fairly sharp
dropoff above $\sim 1 \hmpc$ and goes below 0 at $\gtrsim 3 \hmpc$ (which is
where we stop plotting $\log \xi$).   

The middle subpanel shows the ratios of the cross-correlations of the
subsamples to the full sample auto-correlation and highlights the dependence on
environment.  These differences arise due to the dependence of clustering on 
large-scale environment, the fact that halos in dense environments are more
clustered than halos of the same mass in underdense environments.
We stress that this 
dependence by itself is {\it not} what is commonly referred to as galaxy
assembly bias. To illustrate this we also plot the correlation functions 
of the shuffled galaxies, where galaxies are randomly assigned 
to halos of the same mass, which eliminates any connection to the assembly 
history of the halos.  These correlation functions (the dashed lines) show
essentially the same trends with environment. This is in agreement with 
the conclusions of \citet{abbas05,abbas06} who demonstrate that for the
most part the clustering dependence on large-scale environment can be
explained without resorting to occupancy variation.

The differences between the solid and dashed lines (in the top and middle 
subpanels of Fig.~\ref{Fig:CF_Err}) are the ones reflecting galaxy
assembly bias. Namely, the occupancy variation we quantified in the HOD
coupled with halo assembly bias give rise to these systematic differences 
in the clustering.  We illustrate these in detail in the bottom
subpanel which plots the ratio of original to shuffled correlation functions.
We find that the HOD differences with environment do induce significant 
differences in the large-scale clustering, where for the full sample the
clustering of galaxies is stronger by about $15\%$ than the clustering of
the shuffled galaxies (with no occupancy variation and thus no galaxy
assembly bias). This translates to a $\sim 7\%$ change in the galaxy bias,
in good agreement with the predictions of \citet{croton07}.
The subsample of galaxies in the most dense environments 
exhibits a similar trend, while the galaxies in the least dense regions
are significantly {\it less} clustered than their shuffled counterparts
over most of the range shown.  We discuss in \S~\ref{SubSec:origin} possible 
reasons for this difference.

The bottom part of Fig.~\ref{Fig:CF_Err} investigates the dependence of
clustering on halo age. We see that galaxies in the earliest-formed halos 
are more clustered on large scales than galaxies in the latest-forming halos. 
The clustering differences in this case are much smaller than the differences
with large-scale environment and have similar shapes. This can be readily
seen in the top and middle subpanels, and we remind the reader that
the y-axis in the middle subpanel for halo age spans a much smaller range
than the corresponding one for environment.  Again, we note that while these
differences are directly due to age-dependent halo clustering,
namely reflecting halo assembly bias, these trends only marginally depend
on the occupancy variation with age, as seen by the relatively-small 
differences on large scales between the solid and dashed lines in the middle 
subpanel.

The small-scale clustering in this case shows bigger differences between
the solid and dashed lines (that are noticeable in all subpanels). This
arises mostly because of the large differences in the satellite 
occupation functions, as seen in bottom panel of Fig.~\ref{Fig:HOD_Err}.    
There are relatively more satellites in late-formed halos versus early-formed
halos, especially at the lower mass end of halos that host satellites,
likely simply due to having more time for the satellites to be destroyed
in the early forming halos. This leads to a reversal of the clustering
trend (most notable in the middle panel at $r \sim 1 \hmpc$) and a
stronger small-scale clustering for the galaxies in the most-recently forming
halos.  Interestingly, this is then a case where there is no halo assembly 
bias but the galaxy clustering is still different due to occupancy variation
with halo age. This feature is evident only for the clustering as a function
of halo age and is negligible for the dependence on environment, since in
the latter case the differences between the satellite occupations are 
minuscule (as we have seen in the top panel of of Fig.~\ref{Fig:HOD_Err}).

The bottom subpanel focuses on the galaxy clustering differences due to
the occupancy variations by showing the ratio of the correlation functions
of the original galaxy samples to the shuffled ones. On small scales, we
can see the strong signature of the satellite occupancy variation that we
had just discussed (differences between blue and red lines in the 1-halo
regime below $r \sim 1 \hmpc$).  The ratio for the full sample (black line)
on these scales remains unity, since the shuffling doesn't alter at all
the 1-halo contribution in that case. On larger scales, in the 2-halo regime, 
we see that all three samples exhibit a similar galaxy assembly bias trend, 
where the clustering of galaxies is stronger than that of the shuffled 
galaxies.  We discuss this further below. Appendix~\ref{Sec:AutoCorr} 
presents the auto-correlation functions for these subsets of galaxies 
(instead of the cross correlation with the full sample).

\vspace{0.1cm}
\subsection{Origin of galaxy assembly bias trends}
\label{SubSec:origin}

We note an additional subtle feature in the middle subpanel of the bottom
part of Fig.~\ref{Fig:CF_Err}. As we explained above, the large-scale 
clustering of galaxies as a function of halo age is dominated by the fact 
that, at fixed mass, early-formed halos cluster more strongly than 
late-forming halos (i.e., halo assembly bias). This is reflected by
the difference between the dashed red and blue lines.  When including
the occupancy variation (red and blue solid lines), we see that it acts
to slightly {\it decrease} the clustering differences between early- and 
late-formed halos, for separations larger than $\sim 5 \hmpc$.

These small differences can be understood by examining the changes to the
HOD (as shown in the bottom panel of of Fig.~\ref{Fig:HOD_Err}). The 
central galaxy occupation function for the late-forming halos is shifted
toward higher halo masses, which acts to slightly increase their clustering
(the blue solid line being above the blue dashed line in the middle subpanel
corresponding to formation time). Conversely, the central galaxy occupation
function for the early-formed halos is shifted toward lower halo masses,
resulting in a slightly reduced clustering (the red solid line lying slightly
below the red dashed line).
The interpretation gets a bit more complicated since an opposite trend is seen
for the satellites,  however, we have calculated the effective bias
corresponding to these varying HODs and confirmed that the net effect is
as described above.
We further corroborate this origin of the trend by examining the clustering 
of the different samples when considering only the central galaxies while 
excluding the satellites.

We now turn back to the galaxy assembly bias trends shown on the bottom
subpanels in Fig.~\ref{Fig:CF_Err}, for both halo age and environment,
where we plotted the clustering of the different samples compared to the
clustering of shuffled galaxies where the occupancy variation was erased.
When comparing the galaxy assembly trends for halo age and environment,
we find a similar effect for the galaxies in early-formed halos and for
the galaxies in dense environment (the red solid lines in the two bottom
subpanels).  The clustering difference is in the same sense for galaxies 
in late-forming halos (blue solid line in the bottom subpanel of the bottom
part of the figure). However, for galaxies in dense environments, this
trend reverses, with a weaker clustering than that of the shuffled sample
(blue solid line in the bottom subpanel of the top part of the figure).
We attempt to obtain some insight here regarding the origin of these trends.

We first consider the trends with regard to halo age. In that case, 
we see that regardless of sample used (full sample, early forming, or late 
forming) the galaxy assembly bias trends go in the same direction, with the 
clustering of galaxies stronger than that of the shuffled samples.   We can 
understand why that is by examining the variations in the central galaxies 
HOD in Fig.~\ref{Fig:HOD_Err} and the systematic dependence on halo age in the 
central galaxies SMHM relation (Fig.~\ref{Fig:SMHM}).
For any halo mass, we see that central galaxies tend to occupy first 
the earlier-formed halos.
Thus, it is always the case -- for any range of halo ages -- that the 
relatively earlier-formed halos will be preferentially populated.  
Coupling this with halo assembly bias, namely the stronger clustering
of early-formed halos (for fixed halo mass), results in these galaxies
being more clustered than one would expect otherwise.  Therefore, the 
clustering of {\it any} such sample would always be stronger than the 
clustering of the corresponding shuffled sample.

The variations in the magnitude of the effect can be explained by looking
at the role of the occupancy variation for the satellite galaxies.
For the galaxies in the $20\%$ earliest-formed halos (the blue solid line)
this effect is even more prominent in fact, since these halos also have 
relatively more satellites (compared to the same halos in shuffled case), which
acts to increase the clustering further (via central-satellite galaxy pairs
that contribute as well to the large-scale clustering).  On the other hand,
the $20\%$ latest-forming halos have relatively less satellites (compared
to the same halos in the shuffled case or relative to the HOD for the full 
sample), and this acts to decrease the clustering and slightly reduce the
amplitude of the effect.

For the large-scale environment, we similarly find preferential formation of 
central galaxies in dense environments. This again implies that halos in
relatively denser environments tend to be more populated,  and as these 
halos are more strongly clustered, the galaxies in such halos end up being
more clustered (compared to the shuffled galaxies). 
However, the central galaxies occupancy variation is more nuanced  
for environment than for halo age  and the satellites trend is different 
for them, so the interpretation is more complex.  In particular, it is
non-trivial to explain the sense of the galaxy assembly bias for the
underdense regions, where the clustering is weaker than for the corresponding
shuffled sample. From its occupancy variation, it appears that there are
less satellites overall in that case (for all halo masses), which is possibly
the origin of the reduced clustering we measure.

To confirm this explanation we calculate the effective bias for the 
different HOD variants, by integrating over the halo mass function 
weighted by the different occupation function. For the HOD of the
underdense regions, we find a reduced effective bias compared to that
of the full (or shuffled) sample. When calculating this separately for
the centrals and satellites occupations,  we find that the centrals
term increases a little while the satellites term decreases, in agreement
with the overall effect, leading to the reduced clustering.

We have also examined the clustering results for the L12 model, for which 
there is no change in the satellites occupation function with environment
(Fig.~\ref{Fig:HOD_L12}), to aid our understanding of the role the 
satellites play here. 
The magnitude of the galaxy assembly bias for the underdense regions in
that case is significantly smaller (but still in the same sense 
as for G11).  For lower number density samples (corresponding to more 
massive galaxies), we find that the galaxy assembly bias even switches sign, 
i.e., the L12 cross-correlation function for the underdense regions is 
more strongly clustered than the shuffled case (while the trend remains 
unchanged for G11). This further indicates that the trend observed for G11 
for the underdense regions is due to the satellite occupancy variation.

Finally, we also measured the clustering of the G11 subsamples, when 
considering only central galaxies. We find similar galaxy assembly bias 
trends for the full sample and the dense regions, and a much reduced effect 
for the underdense regions. This again supports our understanding that the
satellites occupancy variation is the main cause of the unique galaxy 
assembly bias we see for the underdense regions, while the central 
occupancy variation is the dominant factor in all other cases.
This difference may also be related to the general shape of the 
cross-correlation function for the underdense regions (induced by the
large smoothing window) and the relative bias factor becoming
negative on large scales (when the cross-correlation function goes 
below zero).

\vspace{0.2cm}
\section{Summary and conclusion} 
\label{Sec:summary}

We have utilized semi-analytic models applied to the Millennium simulation 
to study the occupancy variation leading to galaxy assembly bias. 
We studied in detail the explicit dependence of the halo occupation functions on
large-scale environment, defined as the $5 \hmpc$ Gaussian smoothed dark
matter density field, and on halo formation time, defined as the redshift
at which the halo gained (for the first time) half of its present-day mass.
While related, these two halo properties have only a very loose relation
between them, and probe a different distribution of halos.
We focus our analysis on the $20\%$ subsets of halos at the extremes of the
distributions of each property,  defined separately for each halo mass bin,
so as to eliminate the dependence of the halo mass function on these properties.
We then study the different occupation functions of the galaxies in these
halos and investigate the origin of the variations. We stress that in all 
analyses done here the HODs are calculated directly from the SAM galaxy 
catalogs, and not inferred from clustering measurements. Our key results are 
shown in Figures~\ref{Fig:HOD_Err}, \ref{Fig:SMHM} and \ref{Fig:CF_Err}.

For the dependence on environment, we find small but distinct differences in
the HOD, especially at the ``knee'' of the central occupation function. The 
central galaxies in dense environments start populating lower mass halos, and 
conversely, central galaxies in underdense environments are more likely to
be hosted by more massive halos.  This trend is robust among the two SAMs
we studied. For one of the SAMs (G11), the satellite occupation function
shows similar trends as the centrals, while the other (L12) does not exhibit 
variation for the satellites. We quantify the occupancy variations in terms 
of the changes to the HOD parameters. 

When studying the dependence on halo formation redshift (halo age), we 
find similar but significantly stronger trends for the central occupation 
functions. The satellite occupation function shows a reverse trend, where
the early-formed halos tend to have much fewer satellites than late-formed
halos. This likely arises from having more time for the satellites to merge
with the central galaxy in the older halos.  The relatively stronger trends 
for halo age suggest that this is the more fundamental property (among the 
two) giving rise to galaxy assembly bias.

We gain insight regarding the origin of the central galaxies occupancy 
variation by examining the scatter in the stellar mass -- halo mass relation 
for them and its correlation with halo age and environment.  We find that,
at fixed halo mass, central galaxies in early formed halos or dense
environments tend to be more massive. This directly leads to the occupancy
variation we observe, as for any stellar-mass limited sample, the more 
massive central galaxies will be ``picked'' first in lower-mass halos. 
The dependence on halo age is very distinct, while the dependence on environment
shows more scatter, giving rise to the stronger trends with halo age. 
The trends with halo age can be easily explained as the galaxies in the 
early-formed halos have more time to accrete and form more stars.
These correlations, and the resulting occupancy variation, thus depend on
the specific details of the galaxy formation model.  This direct link to
the correlated nature of the stellar mass -- halo mass relation also
has important implications for models which use subhalo abundance matching 
to connect galaxies with their host halos.

We also examine the auto-correlation and cross-correlation functions of these
different samples and the impact of these occupancy variations, by comparing 
to the clustering of shuffled galaxy samples where the occupancy variation has 
been erased.
We demonstrate the stronger clustering signal of galaxies in the most
dense regions versus least dense regions, and similarly the strong clustering
of galaxies in early-formed versus late-formed halos.  We clarify that while
these clustering differences arise from the dependence of halo clustering
on halo age and environment, they are only marginally affected by the
occupancy variations (and are not, for the most part, what we refer to as 
galaxy assembly bias).

For all samples defined by halo age, the clustering of galaxies is stronger 
than that of the shuffled samples. Namely, we see that the occupation 
variation coupled with halo assembly bias act to increase the 
clustering of galaxies. This effect is explained and dominated by the 
central galaxy occupancy variation. For any range of halo
ages, the earlier-formed halos are preferentially populated. Since these
halos are more strongly clustered, the net effect is a stronger clustering
of the galaxies.  The satellite occupancy variation further modulates
this effect, but is secondary here.  
The same behavior is found for the full sample of galaxies and for galaxies 
in dense environments. This again is due to the tendency to populate more
the halos in denser environments, which are in turn more strongly clustered. 
The galaxies in the most underdense regions exhibit the opposite trend, 
however, with weaker clustering than for the corresponding shuffled sample.
This trend is less intuitive, but is likely caused by the satellites
occupancy variation, as discussed.

Our approach here has already provided considerable insight with regard
to the nature and origin of this complex phenomena. A companion paper 
(Contreras et al., in preparation) is studying the redshift dependence of 
the occupancy variation and galaxy assembly bias in the SAMs, which provides 
a comprehensive view of the evolution of the different trends.
We are also investigating the environment and age occupancy variation
in the EAGLE and Illustris cosmological hydrodynamical simulations 
(Artale et al., in preparation).  

Our study can inform theoretical models of assembly bias as well as 
attempts to determine it in observational data. Additionally, it can 
facilitate creating mock catalogs that incorporate this effect, to
can aid in preparation for future surveys and in evaluating the impact of 
assembly bias on cosmological analyses.
It remains an open (and hotly-debated) question as to what is the level 
of assembly bias in the real Universe. We are hopeful that this work can 
set the stage to developing and applying a method that will conclusively 
determine that.

\smallskip
\acknowledgments
This work was made possible by the efforts of Gerard Lemson and
colleagues at the German Astronomical Virtual Observatory in setting
up the Millennium Simulation database in Garching, and John
Helly and Lydia Heck in setting up the mirror at Durham. We thank Celeste
Artale, Shaun Cole, Ravi Sheth and Zheng Zheng for useful discussions.
IZ and SC acknowledge the hospitality of the ICC at Durham and the helpful 
conversations with many of its members. IZ and NJS acknowledge support by 
NSF grant AST-1612085. IZ was further supported by a CWRU Faculty Seed Grant.
PN \& IZ acknowledge support from the European Research Council, through the
ERC Starting Grant DEGAS-259586. We acknowledge support from the European 
Commission’s Framework Programme 7, through the Marie Curie International 
Research Staff Exchange Scheme LACEGAL (PIRSES-GA-2010-269264) and from a 
STFC/Newton Fund award (ST/M007995/1-DPI20140114). SC further acknowledges 
support the from CONICYT Doctoral Fellowship Programme. NP is supported by 
``Centro de Astronomıa y Tecnologıas Afines'' BASAL PFB-06 and by Fondecyt 
Regular 1150300. CMB \& PN additionally acknowledge the support of the 
Science and Technology Facilities Council (ST/L00075X/1). PN further 
acknowledges the support of the Royal Society through the 
award of a University Research Fellowship. 
The calculations for this paper were performed on the ICC Cosmology
Machine, which is part of the DiRAC-2 Facility jointly funded
by STFC, the Large Facilities Capital Fund of BIS, and Durham
University and on the Geryon computer at the Center for 
Astro-Engineering UC, part of the BASAL PFB-06, which received additional
funding from QUIMAL 130008 and Fondequip AIC-57 for upgrades.

\appendix

\section{A. Sample cuts}
\label{Sec:Cuts}

As described in \S~\ref{SubSec:samples}, we define our density and halo age
samples by ranking the halos according to the property at hand in narrow
bins of halo mass and then selecting the two $20\%$ extremes in each bin.
Our cuts therefore directly depend on halo mass.  The motivation for this
procedure is to is to remove the dependence of the halo mass function on
these properties, effectively ensuring that (for either halo age or 
environment) the two $20\%$ samples have the same halo mass functions. 
This allows for a cleaner comparison of the different HODs, using halos 
of nearly equal mass for the two extremes. 

\begin{figure}
\centering
\includegraphics[width=0.8\textwidth]{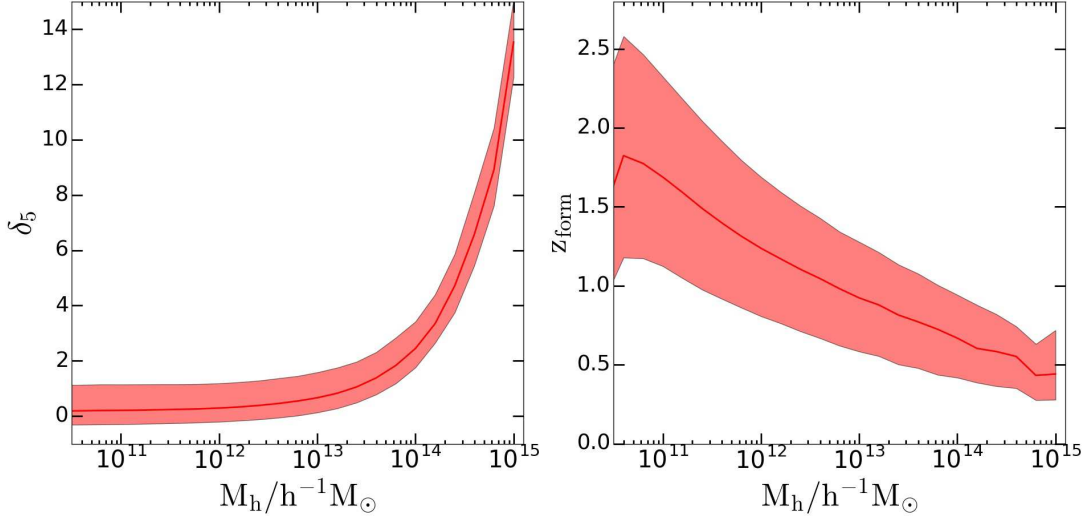}
\caption{\label{Fig:Cuts}
The halo-mass dependent cuts used to define our samples.
The left panel shows how our environment measure $\delta_5$
varies with halo mass. The solid line is the median value of the density 
for each halo mass bin and the shaded region represents the $20\%$ -- $80\%$ 
range of its distribution. The right panel shows the same but now for
halo age as a function of halo mass.}
\end{figure}

Fig~\ref{Fig:Cuts} shows how our environment and halo age sample cuts
vary with halo mass.  The left panel in Fig.~11 shows the environment 
dependence on halo mass, reflecting the known fact that massive halos 
tend to reside in dense environments. The middle solid line is the median value
as a function of halo mass and the shaded region is the $20\%$ -- $80\%$
range of the distribution. This demonstrates the behavior of our halo-mass 
dependent density cut, which adjusts for that exhibiting an upturn toward 
denser environments with increasing halo mass.  The curves that bound 
the shaded regions are exactly the dividing lines defining our different 
environments, i.e, halos that lie above the top curve belong to the
most dense $20\%$ and the halos below the bottom curve make up the 
least dense $20\%$.
The right panel demonstrates how the formation redshift correlates with halo 
mass, exhibiting the known trend that more massive halos tend to form later. 
This, again, is explicitly adjusted for by varying the formation redshift 
boundaries as a function of halo mass, as illustrated.

We note that, in practice, the occupancy variation trends are in fact very 
similar whether one follows this procedure or not. We have verified this by 
defining the $20\%$ extremes samples by a global cut in density, independent 
of halo mass (corresponding to horizontal lines in Fig.~\ref{Fig:Cuts}) and
repeating our analysis.  The clustering of the samples are different in this 
case, since, for example, the most massive halos would predominantly be
included in our densest environment sample and not be represented in the 
underdense sample. This, however, has little impact on the occupancy 
variation, which probes the dependence of the HOD on secondary properties 
other than halo mass. The main difference in the HODs is that they cannot 
all be calculated over the same halo mass range in that case.

\section{B. Auto-correlation functions}
\label{Sec:AutoCorr}

\begin{figure}
\centering
\includegraphics[width=0.48\textwidth]{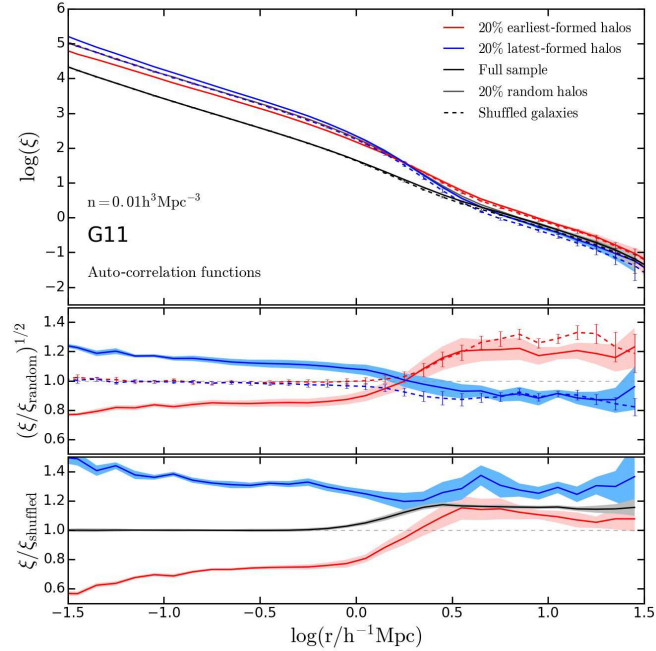}
\caption{\label{Fig:Auto}
Auto-correlation functions for the G11 $n=10^{-2} \hmpcc$ sample, for 
galaxies selected by halo age.  This plot is analogous to the bottom
part of Fig.~\ref{Fig:CF_Err}, but now plotting auto-correlation functions
for all samples, instead of the cross-correlation functions of the 
full sample with the galaxies in early/late-forming halos. Dashed lines are 
the corresponding correlation functions of the shuffled galaxies, as before.
In the top panel we also add the auto-correlation of galaxies in a random 
$20\%$ subset of the halos (grey solid line; see text).
The middle panel shows the correlation functions divided out by the 
auto-correlation function of this random-halos sample. For consistency with
the inferred galaxy bias factors, we plot here the square root of this ratio.
The bottom panel shows the ratio between the correlation functions of the 
original and shuffled samples, for the three main cases. 
The shaded regions represent the uncertainties estimated from 10 jackknife 
realizations.
}
\end{figure}

In \S~\ref{SubSec:clustering}, we analyze in detail the clustering results
of the different samples, using the cross-correlation functions between
the full galaxy sample and the different subsamples.  Here we present the
corresponding results for the auto-correlation samples and explain our
motivation for preferring one over the other.

Figure~\ref{Fig:Auto} illustrates these result for the galaxy samples 
selected according to halo age.  We find a distinct difference in the 
shape of the auto-correlation functions for galaxies in the early- and 
late-forming halos (red and blue solid lines, respectively, in the
top panel), where both 
exhibit a stronger clustering signal than the auto-correlation of the full
sample (black solid line) on scales smaller than $\sim 1 \hmpc$. This 
excess clustering gradually diminishes for scales larger than that, over
roughly the $1$--$7 \hmpc$ range. While perhaps surprising at first 
glance,  this feature arises due to the way the samples are created.
In all such subsamples, $20\%$ of the halos are chosen according to a
specific halo property, and then {\it all} the galaxies in these halos
are included in the sample.  This is systematically a very different 
selection than if we chose $\sim20\%$ of the galaxies without enforcing
the inclusion of all galaxies in a given halo. It acts to ``artificially'' 
increase the small-scale clustering of galaxies mostly due to
the satellite-satellite pairs contribution to the 1-halo term in the
galaxy auto-correlation function.  This excess clustering naturally
decreases when going to larger scales than the size of the halos.

To confirm this explanation,  we study another galaxy subsample in which 
we choose $20\%$ of the halos completely at random and then compute the
auto-correlation function of all galaxies belonging to these randomly-chosen
halos (shown as the solid  grey line in the top panel). We see that this
sample also exhibits this excess clustering on small scales, lying in 
between the auto-correlation functions of the two other subsamples, and 
then converges with the auto-correlation function of the full sample on 
large scales. 
We also performed the simple test of choosing $20\%$ of the galaxies
completely at random, irrespective of which halos they belong to. In that
case (not included in Fig.~\ref{Fig:Auto}), we get the expected result that
this random subset has identical clustering to that of the full sample.

This behavior of the auto-correlation functions, while completely understood, 
is the main reason we prefer to showcase the cross-correlation results in 
the main part of the paper.  The auto-correlation functions for the samples 
defined by environment (not shown here) exhibit the same behavior.
Other than this feature, the rest of the trends are similar whether one 
studies them with the auto-correlation functions or the cross-correlation 
functions, as can be seen in the middle and bottom panels.  One just has 
to be careful to account for the different bias factors that come into
the ratios in each case.
The auto-correlation function measurements are also slightly noisier, 
which is also reflected in the larger errorbars, due to the fact that the 
number of pairs in this case is smaller by about a factor of 5. This is the 
other reason we chose to focus on the cross-correlation results.

In the middle panel of Fig.~\ref{Fig:Auto}, we examine the ratio between
the clustering of galaxies in the $20\%$ early- and late-forming halos
and that of the galaxies in the random $20\%$ of halos (to eliminate the impact
of the small-scale feature). We plot the square root of this ratio since 
two factors of the relative bias parameter come in the auto-correlation 
functions ratio, while only one factor is included in the ratio of the
cross-correlation function and the full-sample auto-correlation function.
It is reassuring to see that the trends on large scales are very similar
qualitatively and quantitatively.  
The trends on small scale are also very similar, just with a slight difference 
in amplitude, likely due to dividing by a larger clustering amplitude.

The results for the galaxy assembly bias shown in the bottom panel of
Fig.~\ref{Fig:Auto}
are also very similar to the ones obtained from the cross-correlation
functions in Fig.~\ref{Fig:CF_Err}. When comparing the values in detail, one 
needs to account for the specific ``assembly bias factors'' (defined in an
analogous manner to galaxy bias factors) in each case. E.g.,
for the galaxies in late-forming halos,  in the auto-correlation functions
ratio the assembly bias factor for that sample appears twice. In contrast,
in the cross-correlation functions ratio this assembly bias appears only
once but gets multiplied by the assembly bias factor for the full galaxy
sample.  Once that is properly accounted for the impact of galaxy assembly
bias on these two different clustering measures is in excellent agreement.


\end{document}